%% file: main.tex
\documentclass[10pt,journal,compsoc]{IEEEtran}
\newcommand{\ie}{\emph{i.e., }}
\newcommand{\eg}{\emph{e.g., }}

\newcommand{\wrt}{\emph{w.r.t. }}
\newcommand{\cf}{\emph{cf. }}

\usepackage{cite}
\usepackage[numbers,sort&compress]{natbib}
\usepackage{amsmath,amssymb,amsfonts}
\usepackage{algorithmic}
\usepackage{graphicx}
\usepackage{textcomp}
\usepackage{xcolor}
\usepackage{bm}
\usepackage{color}
\usepackage{ragged2e}
\usepackage{setspace}
\usepackage{subfigure}
\usepackage{enumitem}
\usepackage{verbatim}
\usepackage[ruled]{algorithm2e}
\usepackage{multirow}
\usepackage{mathtools}
\usepackage[normalem]{ulem}

\newtheorem{theorem}{Theorem}
\usepackage[export]{adjustbox}
\usepackage{hyperref}

\ifCLASSINFOpdf

\begin{document}

\title{Rethinking Missing Data: Aleatoric Uncertainty-Aware Recommendation}

\author{Chenxu Wang,~*Fuli Feng,~Yang Zhang,~Qifan Wang,~Xunhan Hu,~*Xiangnan He
\IEEEcompsocitemizethanks{\IEEEcompsocthanksitem Chenxu Wang, Fuli Feng, Yang Zhang, Xunhan Hu, Xiangnan He are with University of Science and Technology of China.\protect\\
E-mail: wcx123@mail.ustc.edu.com, fulifeng93@gmail.com,zy2015@ mail.ustc.edu.cn,  cathyhxh@mail.ustc.edu.cn,xiangnanhe@gmail.com
\IEEEcompsocthanksitem Qifan Feng is with FaceBook AI, America.
Email:wqfcr@fb.com}%
 \thanks{*Corresponding author}}


\IEEEtitleabstractindextext{%
\begin{abstract}
Historical interactions are the default choice for recommender model training, which typically exhibit high sparsity, \ie most user-item pairs are unobserved missing data. A standard choice is treating the missing data as negative training samples and estimating interaction likelihood between user-item pairs along with the observed interactions. 
In this way, some potential interactions are inevitably mislabeled during training, which will hurt the model fidelity, hindering the model to recall the mislabeled items, especially the long-tail ones.
In this work, we investigate the mislabeling issue from a new perspective of \textit{aleatoric uncertainty}, which describes the inherent randomness of missing data.
The randomness pushes us to go beyond merely the interaction likelihood and embrace aleatoric uncertainty modeling. Towards this end,
we propose a new \textit{Aleatoric Uncertainty-aware Recommendation} (AUR) framework that consists of a new uncertainty estimator along with a normal recommender model.
According to the theory of aleatoric uncertainty, we derive a new recommendation objective to learn the estimator. 
As the chance of mislabeling reflects the potential of a pair, AUR makes recommendations according to the uncertainty, which is demonstrated to improve the recommendation performance of less popular items without sacrificing the overall performance. We instantiate AUR on three representative recommender models: Matrix Factorization (MF), LightGCN, and VAE from mainstream model architectures.
Extensive results on four real-world datasets validate the effectiveness of AUR \wrt better recommendation results, especially on long-tail items. The source code is released at 
\url{https://github.com/wang975679801/AUR}
\end{abstract}

}

\maketitle
\begin{spacing}{0.948}
\input{intro}
\input{problem}
\input{method0418}
\input{exp}
\input{related}
\input{conclusion}
\ifCLASSOPTIONcaptionsoff
  \newpage
\fi



%


\bibliographystyle{IEEEtran}
\bibliography{new_reference}{}
\vspace{-4mm}
\begin{IEEEbiography}[{\includegraphics[width=1in,height=1.25in,clip,keepaspectratio]{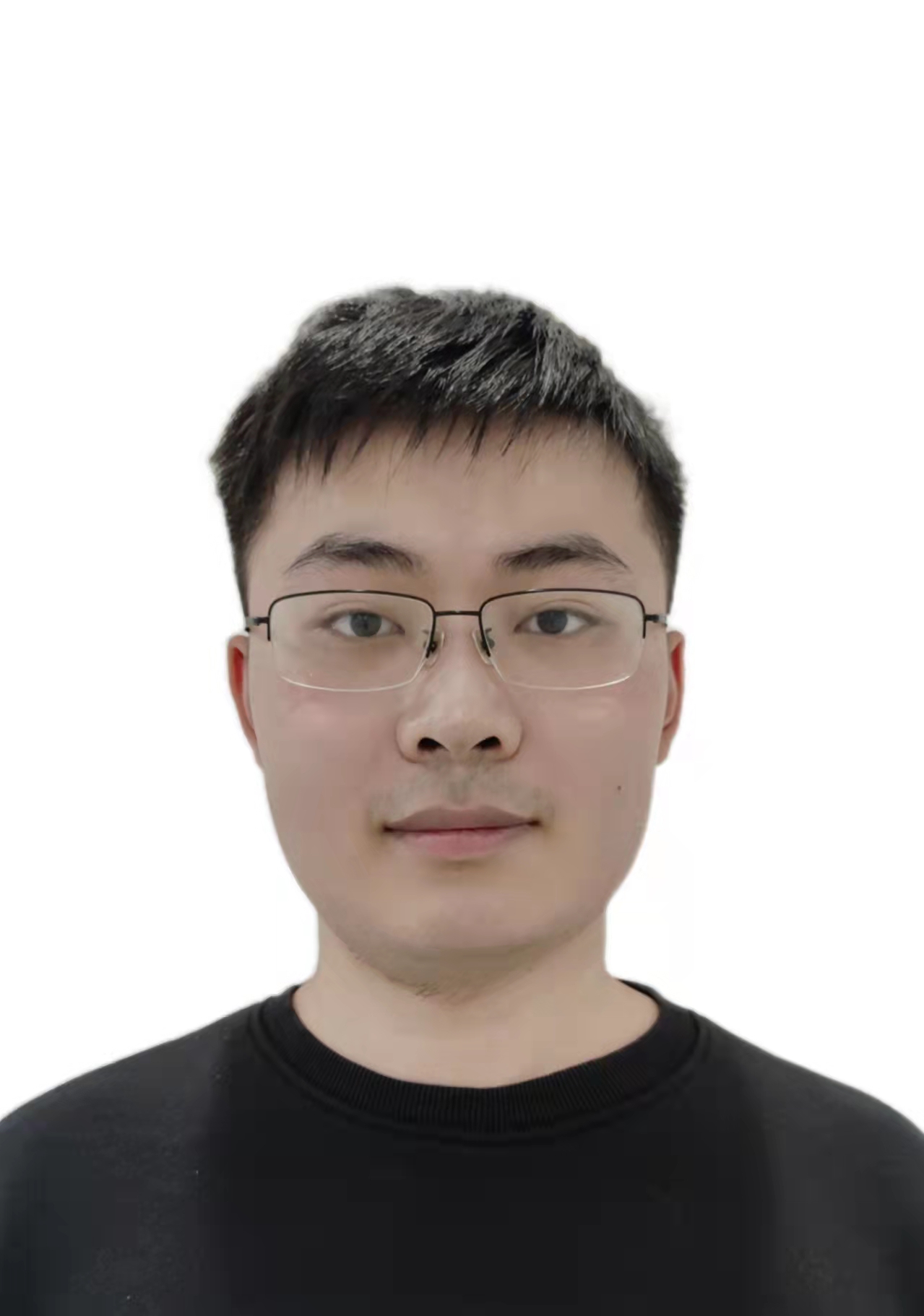}}]{Chenxu Wang}
is currently working toward the MS degree with the College of Big Data, University of Science and Technology of China (USTC), supervised by Prof.Fuli Feng and Prof. Xiangnan He. He received her bachelor's degree from the USTC. His research interest lies in the data mining and recommender system. 
\end{IEEEbiography}
\vspace{-4mm}
\begin{IEEEbiography}[{\includegraphics[width=1in,height=1.25in,clip,keepaspectratio]{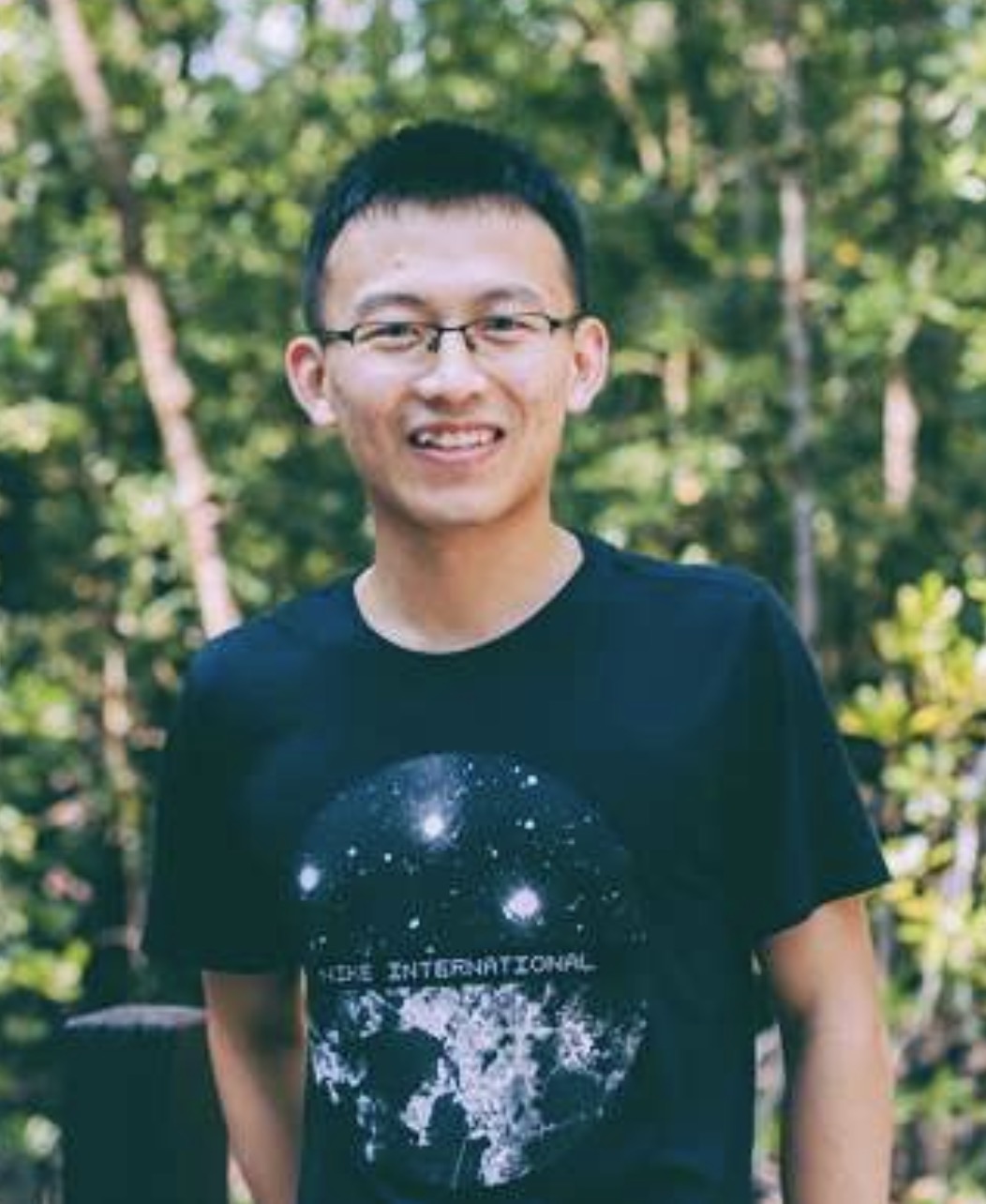}}]{Fuli Feng} 
is a Professor in University of Science and Technology of China. He received B.Eng. from Beihang University and PhD from National University of Singapore. His research interests include information retrieval, data mining, and multi-media analytics. He has 70 publications appeared in several top conferences and journals such as SIGIR, SIGKDD, WWW, TKDE and TOIS. He received the Best Paper Honourable Mention in SIGIR 2021 and Best Poster Award in WWW 2018. Moreover, he has served as the PC member for conferences including SIGIR, WWW, SIGKDD, and the reviewer for journals such as TOIS, TKDE, and Nature Sustainability.
\end{IEEEbiography}
\vspace{-4mm}
\begin{IEEEbiography}[{\includegraphics[width=1in,height=1.25in,clip,keepaspectratio]{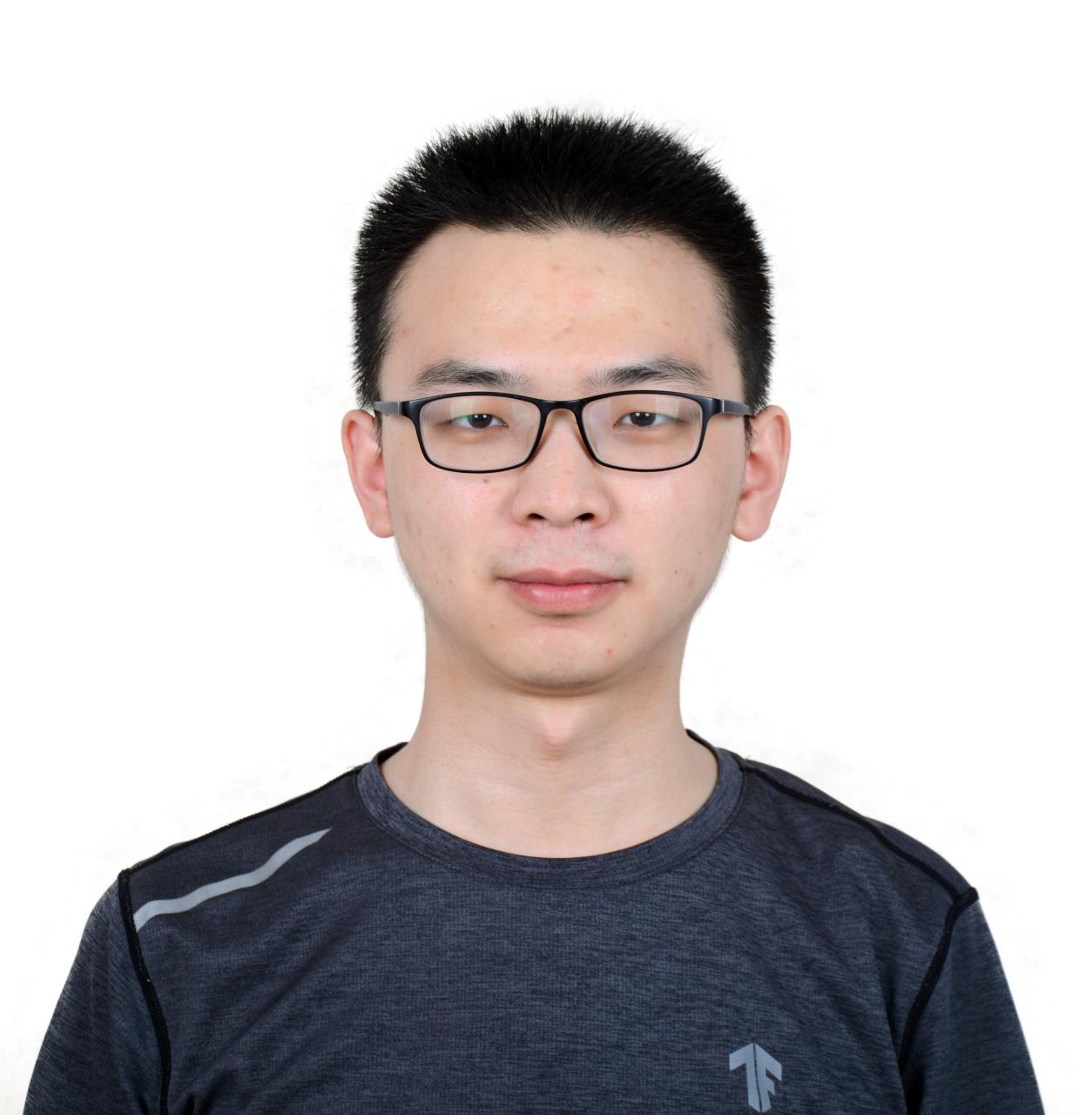}}]{Yang Zhang}
is a Ph.D. student in the School of Information Science and Technology, University of Science and Technology of China (USTC). He received his B.E. degree from the USTC. His research interest lies in the recommender system and causal inference. He has several publications in top conferences/journals. His work on the causal recommendation has received the Best Paper Honorable Mention in SIGIR 2021. He has served as the PC member and reviewer for the top conferences and journals including TOIS, TKDE, TIST, AAAI, WSDM, SIGIR, and SIGKDD.
\end{IEEEbiography}
\vspace{-4mm}
\begin{IEEEbiography}[{\includegraphics[width=1in,height=1.25in,clip,keepaspectratio]{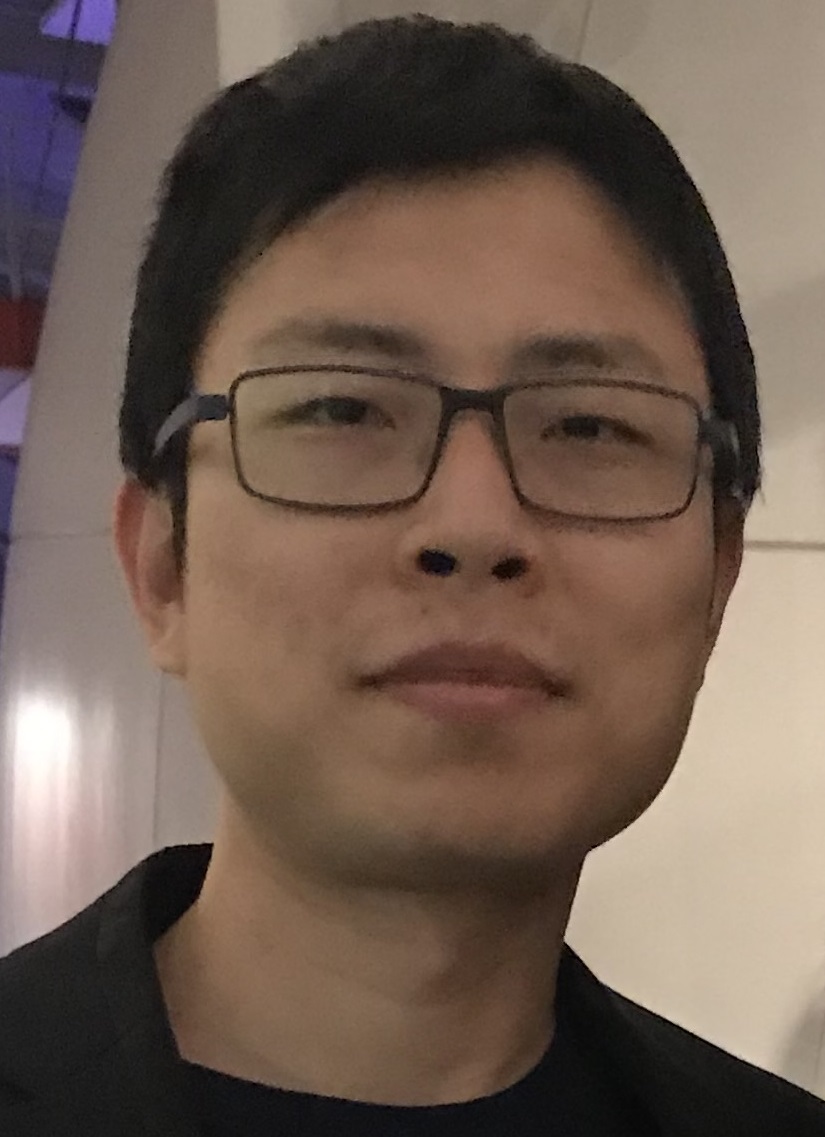}}]{Qifan Wang}
is the Research Scientist at Meta AI, leading a team building innovative Deep Learning and Natural Language Processing models for Recommendation System. Before joining Meta, 
He worked at Google Research, and Intel Labs before joining Meta.
He received his PhD from Purdue University. Prior to that, he obtained both my MS and BS degrees from Tsinghua University. 
His research interests include deep learning, natural language processing, information retrieval, data mining, and computer vision. He has co-authored over 50 publications in top-tier conferences and journals, including NeurIPS, SIGKDD, WWW, SIGIR, AAAI, IJCAI, ACL, EMNLP, WSDM, CIKM, ECCV, TPAMI, TKDE and TOIS. He also serve as area chairs, program committee members, editorial board members, and reviewers for academic conferences and journals.
\end{IEEEbiography}
\vspace{-2mm}
\begin{IEEEbiography}[{\includegraphics[width=1in,height=1.25in,clip,keepaspectratio]{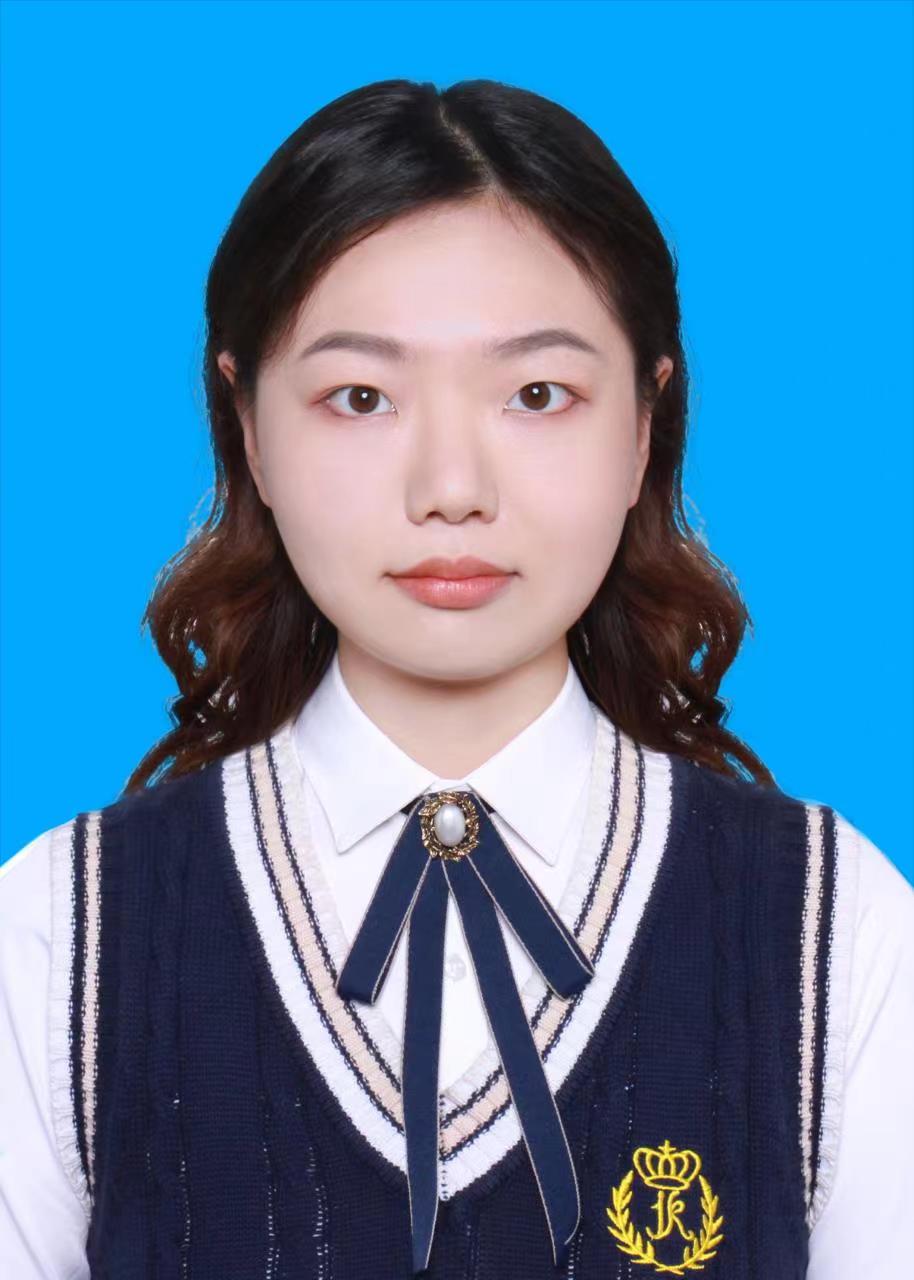}}]{Xunhan Hu}
is currently working toward the MS degree with the College of Big Data, University of Science and Technology of China (USTC), supervised by Prof. Wengang Zhou. She received her bachelor's degree from the USTC. Her research interest lies in the reinforcement learning and multi agent system. 
\end{IEEEbiography}
\vspace{-2mm}
\begin{IEEEbiography}[{\includegraphics[width=1in,height=1.25in,clip,keepaspectratio]{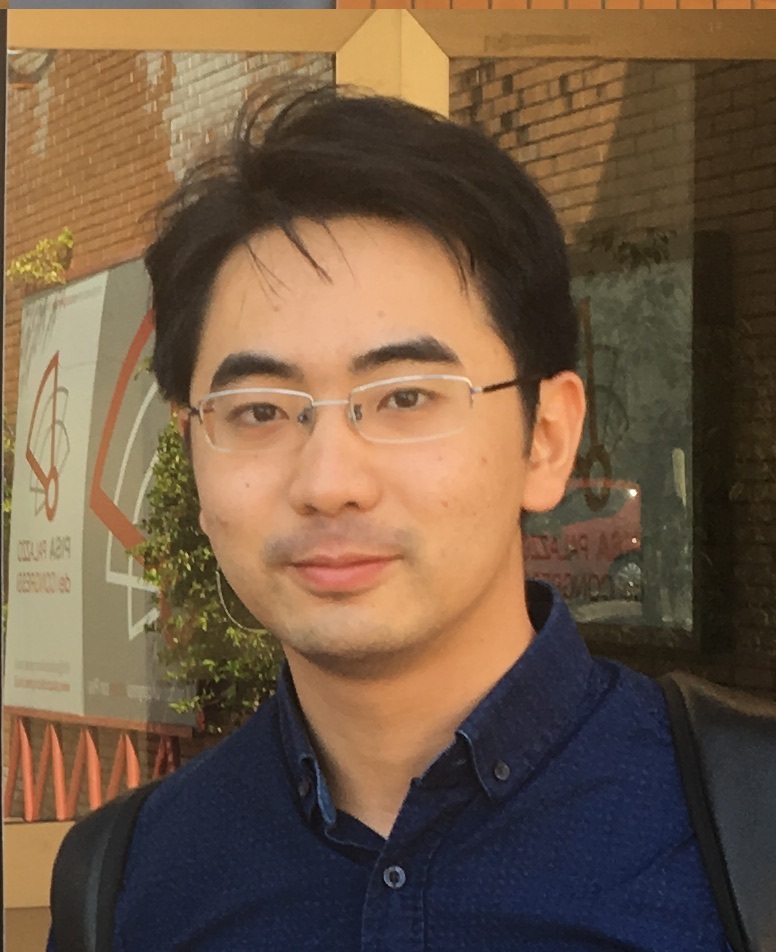}}]{Xiangnan He} 
leads the USTC Lab for Data Science. His research interests span information retrieval, data mining, and multi-media analytics. He has over 100 publications appeared in several top conferences such as SIGIR, WWW, and KDD, and journals including TKDE, TOIS, and TNNLS. His work on recommender systems has received the Best Paper Award Honourable Mention in SIGIR (2021, 2016) and WWW (2018). Moreover, he has served as the Associate Editor for journals including ACM TOIS, IEEE TBD, etc., and (senior) PC member for conferences including SIGIR, WWW, KDD, MM, etc.
\end{IEEEbiography}
%





\end{spacing}
\end{document}

%% file: intro.tex
\IEEEraisesectionheading{\section{Introduction}\label{sec:introduction}}

\IEEEPARstart{R}{ecommender} systems play an irreplaceable role in various online platforms~\cite{news,e-commerce,video}, which aim to facilitate information seeking by providing personalized services.
A canonical paradigm is solving recommendation as a machine learning problem to model the interaction likelihood between user-item pairs for making recommendations. 
A de facto standard is learning the recommender model from historical interactions, which however suffers from severe data sparsity issues~\cite{grvcar2005data}. 
The ratio of missing data, \ie user-item pairs lacking the label of interaction, can reach $99\%$ in many practical cases such as e-commerce~\cite{e-commerce} and social media~\cite{video} due to the huge size of candidate item set which typically increases over time. Worse still, the historical interactions are unevenly distributed over items where long-tail items encounter more missing data, leading to notorious issues like popularity bias~\cite{PDA}.
Therefore, it is essential to properly account for the missing data in recommender training.
%

Existing work typically treats missing data as negative samples~\cite{pan2008one}, making the objective of learning recommender models becoming fitting different labels or expanding the distance between positive and negative samples~\cite{2009BPR}.  
%
The model then produces interaction likelihoods for all negative samples and chooses the top-ranked ones as recommendations.
However, such methods face a mislabeling issue whereby potential positive samples may be incorrectly labeled as negative during the fitting.
Consequently, the model makes inaccurate estimations of interaction likelihoods, which may suppress
potential positive samples, especially the tail ones that have a relatively higher chance to be mislabeled. 
A line of research attempts to tackle this mislabeling issue by sample weighting~\cite{2009Collaborative,2016Modeling,2019Unbiased},
which essentially reduces the weight of potential positive samples in missing data to alleviate their predicted values fitting to wrong labels. 
Nevertheless, these methods will push the model to focus more on popular items, amplifying the popularity bias~\cite{2019Unbiased}.

\begin{figure}[t]
	\centering
	\includegraphics[width=0.98\columnwidth]{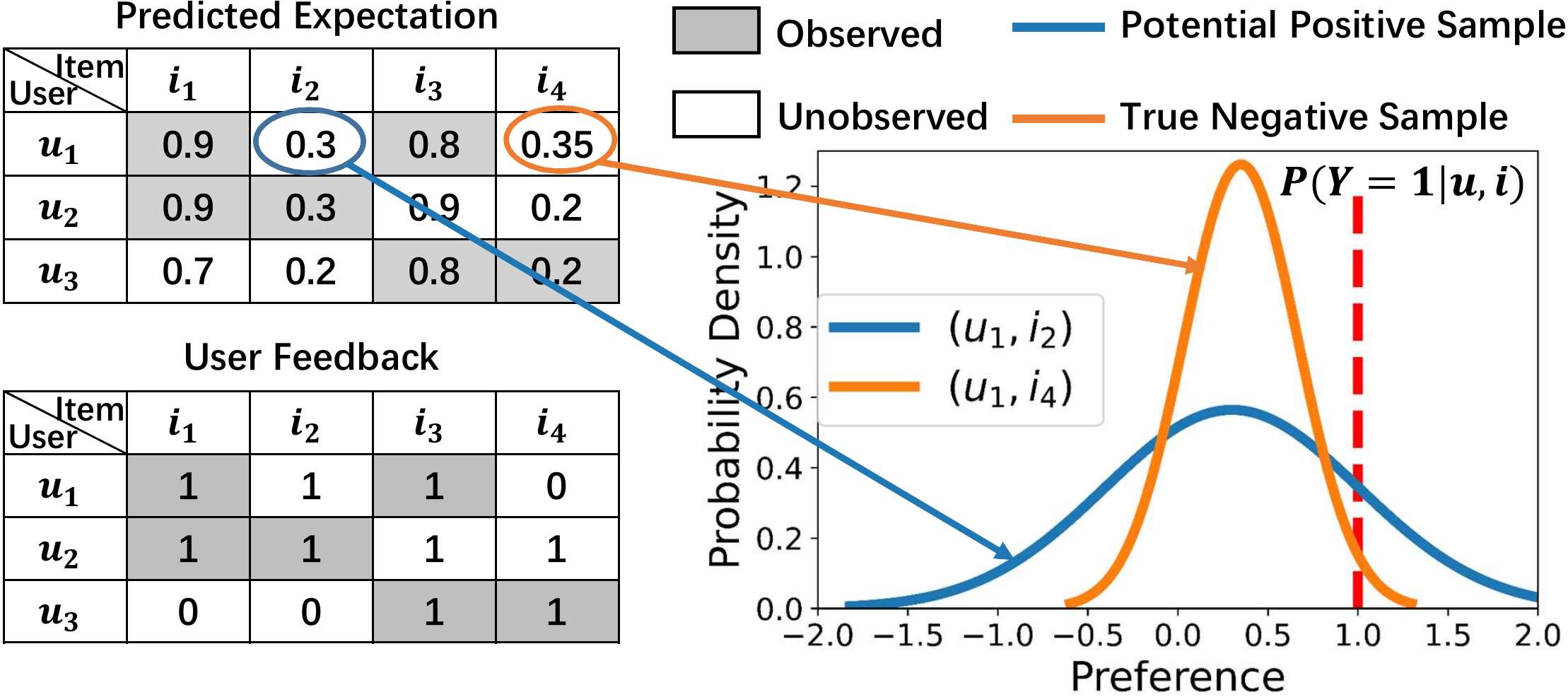}
	\vspace{-5pt}
	\caption{An intuitive example about aleatoric uncertainty in recommendation. We assume $Y_{u_1, i_2} \sim N(0.3, 0.5)$ and $Y_{u_1, i_4} \sim N(0.35,0.1)$. As labels of $i_2$ and $i_4$ have similar predicted expectations, we prefer to recommend $i_2$ that exhibits higher uncertainty.
     }
	\label{intuitive}
	\vspace{-10pt}
\end{figure}

To address the mislabeling issue, we resort to \textit{aleatoric uncertainty}~\cite{2017What} which represents the inherent randomness of the label, \ie a measure of mislabeling.
{Aleatoric uncertainty} has shown success in mitigating the impact of wrong labels in other machine learning problems, such as image segmentation~\cite{zheng2021rectifying} and classification~\cite{uncertainty_app}. 
In this light, we consider incorporating aleatoric uncertainty into recommendation modeling, \ie modeling both the interaction likelihood and aleatoric uncertainty of each user-item pair.
The key lies in accurately estimating
aleatoric uncertainty at the fine-grained level of the user-item pair. 
This is non-trivial due to the specific properties of recommendation data and recommender training that are distinct from the existing research about aleatoric uncertainty in other problems.
For instance, 1) there lacks a formal definition of aleatoric uncertainty in recommendation. 
2) There are complex relations across different samples due to involving the same user or item.
3) The common recommendation losses dissatisfy the assumptions behind the theory of aleatoric uncertainty.

In this work, we propose a new \textit{Aleatoric Uncertainty-aware Recommendation} (AUR) framework. 
In particular, following the representative work PMF~\cite{DBLP:journals/corr/GopalanHB13} and ExpoMF~\cite{2016Modeling}, we assume that the implicit label $Y_{ui}$ obeys a Gaussian Distribution $N(r_{ui},\sigma_{ui}^{2})$.  
Traditional models only estimate the expectation $r_{ui}$, which can be seen as specifying a uniform 
variance for all user-item pairs. Further, the variance $\sigma_{ui}^{2}$ is supposed as the corresponding aleatoric uncertainty for the labels.
As shown in Figure \ref{intuitive}, the potential positive sample $(u_1, i_2)$ with higher uncertainty exhibits a broader Gaussian Distribution than the true negative sample $(u_1, i_4)$.
Since we 
label the missing data as 0
during training, the higher the (learned) variance, the more likely the item will be mislabeled as 0.
It is thus better to give items with higher uncertainty a greater chance of being recommended.
We devise a simple yet effective uncertainty estimator that leverages the idea of collaborative filtering to estimate uncertainty from user and item embeddings, accounting for the relation of uncertainty across different samples. The estimator can be composed of common recommender models that estimate the mean $r_{ui}$ of the distribution.
To learn the parameters of the recommender model and uncertainty estimator, we derive a general training objective to maximize the posterior over both positive and negative samples. 

We argue that aleatoric uncertainty is also an important criteria for making recommendations. 
As items that will be interacted with are mislabeled as negative during training, a higher uncertainty indicates a higher chance that the true label is positive (see Figure \ref{intuitive}).
We thus make recommendations with the linear combination of the expectation and aleatoric uncertainty.
We theoretically and empirically prove that our AUR framework can achieve similar overall performance with the backbone model at least and works particularly well on recommending the tail items.
Moreover, it can alleviate the popularity bias issue since long-tail items will obtain a relatively higher uncertainty.
This is because positive samples on long-tail items typically receive lower predictions of expectation (\ie larger gap to the value of label $1$), pushing the uncertainty estimator to increase its uncertainty prediction.
Consequently, the estimator learns a potential correlation between high uncertainty and low popularity.
Remarkably, the AUR framework is agnostic to the architecture of the recommender model for expectation estimation.
We demonstrate AUR on three representative backbone models: MF~\cite{koren2009matrix}, VAE~\cite{liang2018variational} and LightGCN~\cite{he2020lightgcn} and evaluate it on four real-world datasets, achieving better overall recommendation results and significant improvements on long-tail items.
Our main contributions are as follows:
\begin{enumerate}[leftmargin=*]
\item We present a new perspective of aleatoric uncertainty for recommendation modeling and propose the AUR framework to estimate the uncertainty.
\item We reveal the potential of aleatoric uncertainty in making recommendation and conduct theoretical analysis to demonstrate its superiority over the conventional expectation-based recommendation.
\item We instantiate AUR on three backbone models and conduct extensive experiments, validating the rationality and effectiveness of our method.
\end{enumerate}

%% file: method0418.tex
\section{METHOD}\label{sec:method}

\textbf{Problem Formulation. }
Suppose we have an implicit 
feedback dataset 
$\mathcal{D}$ with $m$ users and $n$ items. The set of users and items are represented as $\mathcal{U}$ and $\mathcal{I}$. 
For any $u \in \mathcal{U}$ and $i \in \mathcal{I}$, the label of pair $(u,i)$ is $Y_{ui}=1$, if $u$ has consumed $i$, and $Y_{ui}=0$ otherwise. We aim to construct a recommender system based on $\mathcal{D}$ to recommend non-consumed items for each user. 


\subsection{Overall Framework}\label{sec:method-overall}
In practical recommender systems, there are many potential positive samples (in $\mathcal{D}$) that are blindly labeled as 0. When training on such data, a certain amount of useful information would be discarded, resulting in an ineffectively trained model.
Therefore, it is an important task to quantify how likely a sample is mislabeled.
While traditional methods treat the recommendation problem as a homoscedastic regression task~\cite{2007Probabilistic}, in this work, we reformulate it as a heterogeneous regression problem where samples have different variances, and propose a \textit{Aleatoric Uncertainty-aware Recommendation} (AUR) framework.


\begin{figure}
\centering	\includegraphics[width=0.8\columnwidth]{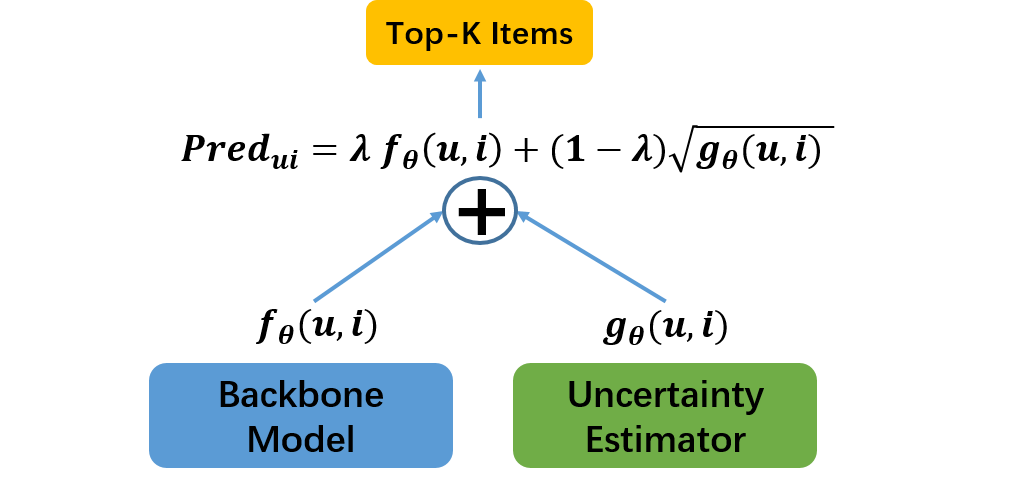}
\caption{
Model overview of our Aleatoric Uncertainty-aware Recommendation framework.
}
	\label{pic_overall}
\end{figure}

\vspace{+5pt}
In our framework, we assume that the implicit feedback label $Y_{ui}$ follows the Gaussian Distribution $N(r_{ui},\sigma_{ui}^2)$, where $r_{ui}$ denotes the mean predicted by a backbone recommender model $f_{\theta}$  parameterized with $\theta$ (\ie $r_{ui}=f_\theta(u,i)$).
The variance $\sigma_{ui}^2$ represents the stability of the Gaussian Distribution, which is a variable learned from the implicit data. 
A large variance $\sigma_{ui}^2$ usually indicates that the given label is unstable,
\ie a label with a large variance is likely to be incorrect.
We thus use the variance of the label to quantify the likelihood of mislabeling, which is referred to as aleatoric uncertainty, and utilize it for enhancing recommendation quality. Notably, the $\sigma_{ui}^2$ is also obtained by an uncertainty estimator $g_{\phi}$  parameterized with $\phi$ (\ie $\sigma_{ui}^2=g_\phi(u,i)$). The core of our framework lies in estimating the mean and variance of the Gaussian distribution with the backbone recommender and the uncertainty estimator and generating recommendations by combining the mean and variance, as shown in Figure~\ref{pic_overall}.

\vspace{+5pt}
\noindent \textbf{$\bullet$Backbone recommender and uncertainty estimator.}
Our framework is model-agnostic. The backbone recommender to estimate the mean $r_{ui}$  can be any kind of collaborative filtering model, such as SVD-Based models~\cite{2009BPR}, Graph-Based models~\cite{he2020lightgcn}, and DNN-Based models~\cite{liang2018variational}. 
As for the design of \textbf{uncertainty estimator}, to ensure the non-negativity of $\sigma_{ui}^2$, we naturally consider denoting it by following the exponential form:
\begin{equation}\small
\sigma_{ui}^2=g_{\phi}(u,i)=\frac{e^{h_{\phi}(u, i)}}{K},
\label{exponential}
\end{equation}
where $K$ controls
the scale of exponential function.
%
Notice that the uncertainty of the label $Y_{ui}$ is a combination of the uncertainty signals from the user and item. 
We learn latent item representation $q_i$ and user representation $p_u$, similar to recommender models, and combine them as $h_{\phi}(u, i)$. 

As to item $i$, we learn a projection $f_{item}:e_i \rightarrow q_i$ to 
mine the latent factors affecting uncertainty,
where $e_i$ denotes the one-hot encoding of item $i$.
For user representation, intuitively, users with similar item interactions should have similar uncertainty.
We employ the combination of all item representations in the user's historical interaction as the user representation. Formally,
$p_u=\sigma(\frac{1}{\sqrt{|\mathcal{H}_u|}}\sum_{i\in \mathcal{H}_u}z_i),$
where $\mathcal{H}_u$ denotes the historical interaction set of user $u$, and $\sigma$
indicates an activation function. $z_i$ is the user history information encoded by another projection. This design is similar to \cite{koren2008factorization}, which validates the rationality of using two different projections. $f_{user}:e_i \rightarrow z_i$.

Then $h_{\phi}(u, i)$ is simply defined as the inner product of item representation $q_i$ and user representation $p_u$, \ie $h_{\phi}(u,i)=<p_u,q_i>$. For briefness, we denote the value of $h_{\phi}(u, i)$ as {$s_{ui}$} in the rest of the paper.

\vspace{+5pt}
\noindent\textbf{$\bullet$Recommendation.}
Once we obtain the aleatoric uncertainty, we will leverage it to adjust the prediction of the backbone recommender by linearly adding it as follows:
\begin{equation}
    Pred_{ui}=\lambda f_{\theta}(u,i)+(1-\lambda) \sqrt{g_{\phi}(u,i)},
\end{equation}
where $Pred_{ui}$ is the final recommendation score and is used to generate $Top-K$ items, and $\lambda$ is used to control the influence of the uncertainty on recommendations. As mentioned above, samples with high uncertainty are more suppressed by mislabeling issue. Therefore, we need to compensate these samples more. 
In mathematics, we use the confidence upper bound of normal distribution to recommend items.

\subsection{Training Procedure}\label{sec:training}
A default choice for estimating the parameters of backbone model and uncertainty estimator is minimizing the negative log-likelihood of training samples. Formally, the negative log-likelihood of $Y_{ui}$ is $-logP(Y_{ui}|r_{ui},\sigma_{ui}^2)=\frac{(r_{ui}-Y_{ui})^2}{2\sigma_{ui}^2}+\frac{1}{2}log \ \sigma_{ui}^2$.
Due to the huge amount of training instances, we usually consider the mini-batch training strategy. 
We adopt the same user-batch strategy used in \cite{liang2018variational}. We first randomly sample a batch of users $\mathcal{U}_{batch}$ from $\mathcal{U}$, then the user batch loss $L(\mathcal{U}_{batch},\mathcal{D};\theta,\phi)$ can be formulated as:
\begin{equation}\small
\begin{split}
\sum_{u\in \mathcal{U}_{batch}, i \in \mathcal{I}}w_{ui}\left(
    \frac{(r_{ui}-Y_{ui})^2}{
        \frac{2}{K}e^{s_{ui}}}+\frac{1}{2} \ s_{ui}
\right)+\Omega,
\label{loss_overall}
\end{split}
\end{equation}
where $w_{ui}$ denotes the weight of different samples in the training set $\mathcal{D}$, and $\Omega$ denotes the regularization term.


Noticing that optimizing the two models simultaneously can lead to unstable issues and overfitting~\cite{yu2020sampler},
we propose to train them sequentially in an Expectation-Maximization like algorithm. In particular, we first train the backbone model as our expectation estimator without considering the uncertainty, and then train the uncertainty estimator while fixing the expectation model. 
Sequential training also simulates the situation where our method is grafted onto an existing model, which is required by industry applications. We also show the results of the joint learning of expectation and uncertainty in the experiments (\cf Table~\ref{variant}).
\begin{algorithm}
  \SetAlgoLined
  \KwData{Hyperparameters $K$,$\alpha$,$\beta$,$\gamma$,$\lambda$,$\mu$}
  \KwResult{Learned model parameters}
  Initialize $\theta$, $\Phi$\;
  //\textbf{Backbone Training Process}\;
  \While{not converge }{
  Compute $r_{ui}=f_{\theta}(u,i)$ for each (u,i)\;
  Update $\theta$ according to Equation \eqref{loss_backbone}\;
  }
  //\textbf{Uncertainty Estimator Training Process}\;
  \While{not converge }{
  Compute $r_{ui}=f_{\theta}(u,i)$ for each (u,i)\;
  Compute  $s_{ui}=h_{\phi}(u,i)$ for each (u,i)\;
  Update parameters $\phi$ by Equation \eqref{loss}\;
  }
  Return the learned $\theta$ and $\phi$\; 

  \caption{Overall Training Process for AUR}
  \label{train_process}
\end{algorithm}

\subsubsection{Backbone Training}\label{expectation_training}
To alleviate the imbalance issue of positive and negative samples, we leverage a sampling strategy.
We assign the weight to each instance $w_{ui}$ as follows:
\begin{equation}\small
    w_{ui}\sim
   \begin{cases}
   1 &\mbox{$Y_{ui}=1$,}\\
   Bernoulli(\mu) &\mbox{$Y_{ui}=0$.}
   \end{cases}
   \label{weight}
\end{equation}
where $\mu$ denotes the sampling rate for negative samples.
The aleatoric uncertainty is not considered for the backbone model,  
$s_{ui}$ is thus fixed for all users and items. 
We set $s_{ui} = 0$ and use $l_2$-norm as regularization, \ie $\Omega=\lambda||\theta||^2$, simplifying the training objective for backbone model as:
\begin{equation}\small
\begin{split}
&L(\mathcal{U}_{batch},\mathcal{D};\theta)=\sum_{u\in \mathcal{U}_{batch}, i \in \mathcal{I}}w_{ui}\left(
    \frac{(r_{ui}-Y_{ui})^2}{2}
\right)+\lambda ||\theta||^2.
\label{loss_backbone}
\end{split}
\end{equation}
The first term is essentially a mean square error, one of the common recommendation losses. 
The form of the loss is decided by the prior assumption of data distribution. Other losses like BPR \cite{2009BPR} corresponds to distributions other than Gaussian Distribution.
We belief Gaussian Distribution is a reasonable assumption since Equation ~\eqref{loss_backbone} can lead to stronger backbone models than BPR (\cf Figure~\ref{fair}).
\subsubsection{Uncertainty Estimator Training}
The uncertainty estimator is trained on top of the backbone model. In this stage, we also need to balance the positive and negative samples, although the learning objective is not the hard label of implicit data. 
To speed up training, we replace the sampling process by directly assigning fixed weight to each instance.
The weight $w_{ui}$ there can be formulated as:
\begin{equation}\small
    w_{ui}=
   \begin{cases}
   \alpha &\mbox{$Y_{ui}=1$,}\\
   1 &\mbox{$Y_{ui}=0$.}
   \end{cases}
   \label{weight}
\end{equation}
where $\alpha$ is a key hyperparameter for training the uncertainty estimator. More specifically, $\alpha$ can be adjusted to control the strength of 
emphasizing the tail items (\cf Figure~\ref{hyper}). 

To optimize the uncertainty estimator, we first derive the distribution of uncertainty $P(s_{ui}|r_{ui},Y_{ui})$. According to Bayes theorem, we can get:
\begin{equation}\small
    P(s_{ui}|r_{ui},Y_{ui})=\frac{P(Y_{ui}|r_{ui},s_{ui})P(r_{ui}|s_{ui})P(s_{ui})}{P(r_{ui},Y_{ui})}.
\end{equation}
Since the expectation and variance of a Gaussian Distribution are independent statistics~\cite{casella2021statistical}, $s_{ui}$ and $r_{ui}$ are independent. We have $P(r_{ui}|s_{ui})=P(r_{ui})$. Note that when training the uncertainty estimator, $r_{ui}$ is fixed and treated as a constant. Then we have $P(s_{ui}|r_{ui},Y_{ui}) \propto P(Y_{ui}|r_{ui},s_{ui})P(s_{ui})$. We assume that $s_{ui}$ has a prior of zero-mean Gaussian Distribution: $s_{ui} \sim N(0,\tau^2)$.
Combined with Equation \eqref{exponential}, we obtain the posterior probability of $s_{ui}$ and minimize its negative logarithm to optimize the parameters of uncertainty estimator. Formally,
\begin{equation}\small
    -logP(s_{ui}|r_{ui},Y_{ui})=\frac{(r_{ui}-Y_{ui})^2}{\frac{2}{K}e^{s_{ui}}}+\frac{1}{2}s_{ui}+\frac{s_{ui}^2}{\tau^2}+c,
\end{equation}
where $c$ denotes a constant. After simplifying the form, the final optimization objective is:
\begin{small}
\begin{equation}
\begin{split}
&L(\mathcal{U}_{batch},\mathcal{D};\phi)=\sum_{u \in \mathcal{U}_{batch},i \in \mathcal{I}}w_{ui}\left(
    \frac{(r_{ui}-Y_{ui})^2}{e^{s_{ui}}}+\beta s_{ui}+ \gamma s_{ui}^2
\right),
\label{loss}
\end{split}
\end{equation}
\end{small}
where $\beta$, $\gamma$ are hyperparameters.
This training process can be understood as finding the best variances such that the data likelihood is maximized.
After obtaining the optimal values of the variance, our model makes recommendations based on both of the expectation and aleatoric uncertainty. The overall training process is organized as Algorithm \ref{train_process}.
\begin{figure}[t]
	\centering
	\subfigure{
		\includegraphics[width=0.44\columnwidth]{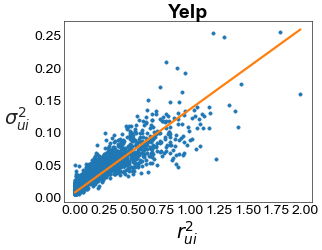}
	}
	\subfigure{
		\includegraphics[width=0.44
		\columnwidth]{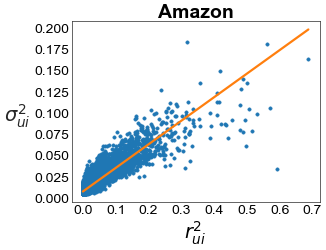}
	}
 	\vspace{-10pt}
	\caption{Relationship between $r_{ui}^2$ and $\sigma_{ui}^2$.}
	\vspace{-10pt}
\label{r_sigma}
\end{figure}

\subsection{Theoretical Analysis}\label{sec:algo}
In this subsection, we first show that the overall performance of the recommendation with aleatoric uncertainty is at least as good as the backbone models using expectation. The statement is further verified empirically with results from four real datasets (\cf Table~\ref{main result}). We then explain the mechanism of improving the tail performance.

\noindent \textbf{$\bullet$ Overall Performance. }
We first state the property that connects the aleatoric uncertainty with the expectation. 
\begin{theorem}
Denote the residual of the expectation as $\Delta_{ui}=r_{ui}-Y_{ui}$. 
For user $u$, denote the distributions of $\Delta_{ui}^2$ and $e^{s_{ui}}/K$ over the items as $p_u$ and $q_u$ respectively.
If the weights $w_{ui}$ are the same for all user-item pairs, the following inequality holds:
\begin{equation}\small
    L(\mathcal{D}) \ge \sum_{u}KL(p_{u}||q_{u}) + c,
\end{equation}
where $L(\mathcal{D})$ is the loss of uncertainty estimator in Equation \eqref{loss}; $KL(p_u||q_u)$ is the KL-Divergence of the two distributions; and $c$ is a constant. 
\end{theorem}
\textbf{Proof of Theorem 1}\label{proof}
The training objective of uncertainty estimator, \ie Equation (\ref{loss}) can be written as:
\begin{equation}
\small
\begin{split}
&L(D)=\sum_u\sum_{i}(\frac{\Delta_{ui}^2}{e^{s_{ui}}}+\beta s_{ui}+ \gamma s_{ui}^2)
\end{split}
\end{equation}
here we assume $w_{ui}$ or $\alpha$ in equation (\ref{weight}) is 1, but it is trivial to obtain a similar inequality for $\alpha \neq 1$.
We obtain the normalized distribution of $p_u$ and $q_u$ as $p_{ui}= \frac{\Delta_{ui}^2}{K_{u}}$ and $q_{ui}=\frac{e^{s_{ui}}}{R_{u}}$,
where $K_{u}=\sum_{i}\Delta_{ui}^2, R_{u}=\sum_{i}e^{s_{ui}}$, and $p_{ui}$ and $q_{ui}$ denote the value of $p_u$ and $q_u$ on the $i$-$th$ item.
According to the common inequality: $lnx+1 \le x (x>0)$, we can derive the inequality:
\begin{equation}
\small
    \frac{\Delta_{ui}^2}{e^{s_{ui}}}=  \frac{\Delta_{ui}^2}{K_{u}}\frac{K_{u}}{e^{s_{ui}}}\ge \frac{\Delta_{ui}^2}{K_{u}}(ln\frac{\frac{\Delta_{ui}^2}{K_{u}}}{\frac{e_{s_{ui}}}{R_{u}}}+ln\frac{K_{u}^2}{R_{u}}+1-ln\Delta_{ui}^2)
\end{equation}
By substituting the above inequality into $L(D)$, we have:
\begin{equation}\small
\begin{split}
&L(\mathcal{D})\ge \sum_{u}\sum_{i}(p_{ui}(ln\frac{p_{ui}}{q_{ui}}-lnR_{u})+\beta s_{ui}+ \gamma s_{ui}^2)+c_1\\
&=\sum_{u}KL(p_{u}||q_{u})+\sum_{u}\sum_{i}(\beta s_{ui}+ \gamma s_{ui}^2 -lnR_{u})+c_1
\end{split}
\label{eq21}
\end{equation}
Assume that $s_{uk}=max_{i}s_{ui}$, we have $lnR_{u}\le ln|\mathcal{I}|e^{s_{uk}}=s_{uk}ln|\mathcal{I}|$ and $\mathcal{I}$ is the set of all items. According to the Extremum Theorem of a quadratic function, we can derive:
\begin{equation}
\begin{split}
&\beta s_{ui}+ \gamma s_{ui}^2 \ge \frac{-\beta ^2}{4\gamma} \ \ \ \ \ \ \  for \ i \neq k\\
&(\beta-ln|\mathcal{I}|) s_{uk}+\gamma s_{uk}^2 \ge \frac{-(\beta-ln|\mathcal{I}|) ^2}{4\gamma} \ for \ i = k
\end{split}
\end{equation}
Therefore, the second term in the RHS of the inequality (\ref{eq21}) is also lower bounded by a constant.
Overall, we can obtain the inequality:
\begin{equation}
\small
    L(\mathcal{D}) \ge \sum_{u}KL(p_{u}||q_{u})+c
\end{equation}
This completes the proof of the theorem$\blacksquare$

The above theorem essentially states that the KL-Divergence between the distributions of expectation and uncertainty is bounded by the loss function that is being optimized. In other words, the lower the loss can be achieved, the closer the two distributions will become. Therefore, the expectation and the uncertainty are highly positively correlated, and thus the overall performance of uncertainty recommendation should be at least on par with the recommendation models using expectation.

To further validate the above statement, we report the empirical results on two real world datasets in Figure \ref{r_sigma}. Specifically, we randomly select one user from each dataset and show the visual relationship between $r_{ui}^2$ and $\sigma_{ui}^2$. 
As shown in Figure \ref{r_sigma}, it is clear that expectation and uncertainty have strong linear positive correlation for both users, which confirms the theorem. The yellow line indicates the best fit line. Moreover, we calculate the mean pearson correlation coefficient on different datasets over the users, where it is 0.69 on Amazon-Book and 0.74 on Yelp (see dataset descriptions in Section \ref{ssec:exp_setting}).

\noindent \textbf{$\bullet$ Tail Performance. }
We also claim that using uncertainty to make recommendations improves the quality of tail items recommendations. 
The reason is that when a positive sample is not able to be retrieved by the expectation $r_{ui}$, its variance $\sigma_{ui}^2$ would be learned to be a large value. In most cases, the expectation of tail positive samples is lower.
This positive sample is likely to be captured by our AUR model. Especially, when the hyperparameter $\alpha$ in Equation \eqref{weight} increases, the $\sigma_{ui}^2$ of tail items would also increase. Therefore, we name $\alpha$ as the \textbf{tail-controlling coefficient}.

We use an example to illustrate how AUR improves tail performance in Table \ref{pop}. We select a user and four items from the Amazon-Book dataset to compare the learned expectations and uncertainties. The backbone model is MF. We count item popularity in the training set while the value of user feedback is retrieved from the testing set. From the table, we can see that $i_4$ has a much lower popularity compared to $i_1$, as well as its expectation.
Despite that item $i_1$ has been interacted many times, it is actually a true negative sample for the user.
Both the popularity and the expectation of $i_4$, a true positive sample, are the lowest among the four items, but it still has the highest uncertainty indicating that $i_4$ is a potential positive sample. Although the tail positive sample $i_4$ has a lower expectation, the uncertainty may be higher. Therefore, AUR would make more recommendations to $i_4$. We provide more analysis in the experiments.
\begin{table}
\small
\centering
\caption{Case study of selected user-item pairs. We show four items with their expectation, uncertainty, and popularity in Amazon to explain how AUR improves the tail performance. }
\vspace{-4mm}
\begin{tabular}{|c|cccc|}
\hline
Items         & $i_1$   & $i_2$   & $i_3$   & $i_4$   \\ \hline
Expectation   & 0.8038 & 0.7784 & 0.7654 & 0.4648 \\
Uncertainty   & 0.3188 & 0.2952 & 0.3050 & 0.3277 \\
Popularity    & 128    & 44     & 59     & 40     \\
User Feedback & 0      & 0      & 0      & 1      \\ \hline
\end{tabular}
\label{pop}
\vspace{-6mm}
\end{table}

\subsection{Discussion}\label{sec:discussion}
\noindent \textbf{$\bullet$ Relevance to Sample Weighting. }
We first discuss the connections and distinctions between our approach and previous methods. 
In the previous discussion, \cite{2009Collaborative,reweight1} heuristically assigned weights $c_{ui}$ to each training sample. In our work, if we fix $s_{ui}=log(\frac{1}{c_{ui}})$ and $\alpha=1$ in Equation \eqref{loss} to optimize the backbone model, our approach is equivalent to \cite{2009Collaborative,reweight1}.
Intuitively, for the positive samples, high correlation between $(u,i)$ indicates low variance \cite{2009Collaborative}. For the negative samples, high correlation indicates that the item $i$ might be a potential positive sample for user $u$, resulting in a high variance \cite{reweight1}.
Therefore, assigning fixed weights to re-weight MSE loss in the training process is essentially to determine the uncertainty of each sample.
Previous methods specify/fix the variance to learn the expectation. In contrast, our approach first learns an expectation, and then estimates the variance based on the learned expectation. During the recommendation stage, previous methods recommend items using expectation while we make recommendations with variance to better capture the uncertainty in the data.

\noindent \textbf{$\bullet$ Relevance to Probabilistic Methods. }
Similar to AUR, some probabilistic models, \textit{e.g.}, PMF \cite{2007Probabilistic} and ExpoMF \cite{2016Modeling}, also assume the implicit labels obey Gaussian Distributions. Differently, they take the variances of Gaussian Distributions as hyper-parameters, and assume all $(u,i)$ pairs have the same variances. 
AUR estimates the variances for each $(u,i)$ pair with the proposed learning strategy, obtaining more accurate uncertainties at a fine-grained user-item pair level.
Moreover, these probabilistic models generate recommendations with the predicted expectations, which may over-recommend popular items,as discussed in Section~\ref{exp_prob}. Fundamentally different to them, AUR makes recommendations based on the uncertainties, aiming at providing more accurate and more balanced recommendations.

%% file: exp.tex
\section{EXPERIMENTS}\label{sec:experiment}
In this section, we conduct experiments on four real-world datasets to answer the following questions:
\textbf{RQ1:} 
How is the performance of AUR compared with expectation-based recommender models, the debiasing methods, and the probabilistic methods?
\textbf{RQ2:} How does each design of AUR affect its effectiveness \wrt the recommendation performance?
\textbf{RQ3:} Do recommendations given by uncertainty exhibit properties different from those given by expectation?

\subsection{Experimental Settings}\label{ssec:exp_setting}

\textbf{Datasets. }
We conduct experiments on four benchmark datasets, namely Yelp2018, Amazon-book, Adressa, and MovieLens-10M, which we abbreviate as Yelp, Amazon, Adressa, and ML-10M, respectively. For Yelp and Amazon, we follow the same setting as LightGCN~\cite{he2020lightgcn} to filter users and items and split each dataset into training and testing sets. For Adressa and ML-10M, we adopt the setting of MACR~\cite{wei2021model} to filter the datasets. To evaluate the performance of tail item recommendation, we define tail items as items with fewer interactions and make sure that they comprise 50\% of all interactions in the corresponding dataset. The remaining items are treated as head items. On Yelp and Amazon, 86.3\% and 84\% of the items were tail items, respectively. As for the Adressa and ML-10M datasets, approximately 85\% of the items were tail items.




\begin{table*}[]
\caption{Performance comparison of top-$K$ recommendation performance between expectation-based models and AUR  on Yelp, Amazon, Adressa, and ML-10M, under the Tail Absolute, Tail Relative, and Overall evaluation protocols. The metric Recall and NDCG are denoted as R and N, respectively.} 
\vspace{-0.3cm}
\small
\resizebox{\linewidth}{50mm}{
\begin{tabular}{|c|c|cccc|cccc|cccc|}
\hline
\multirow{2}{*}{Datasets}     & \multirow{2}{*}{Methods} & \multicolumn{4}{c|}{Tail Absolute}                                                                                 & \multicolumn{4}{c|}{Tail Relative}                                                                                 & \multicolumn{4}{c|}{Overall}                                                                                       \\ \cline{3-14} 
                              &                          & R@20                       & R@50                       & N@20                       & N@50                        & R@20                       & R@50                       & N@20                       & N@50                        & R@20                       & R@50                       & N@20                       & N@50                        \\ \hline
                              & MF                       & 0.0046                     & 0.0180                     & 0.0025                     & 0.0071                      & 0.0522                     & 0.1010                     & 0.0343                     & 0.0498                      & 0.0639                     & 0.1215                     & 0.0531                     & 0.0745                      \\
                              & MF-AUR                   & 0.0184                     & 0.0459                     & 0.0112                     & 0.0203                      & 0.0650                     & 0.1230                     & 0.0430                     & 0.0614                      & 0.0689                     & 0.1316                     & 0.0563                     & 0.0796                      \\ \cline{2-14} 
Yelp                          & LGCN                     & 0.0081                     & 0.0296                     & 0.0044                     & 0.0116                      & 0.0590                     & 0.1126                     & 0.0390                     & 0.0562                      & 0.0673                     & 0.1280                     & 0.0557                     & 0.0782                      \\
                              & LGCN-AUR                 & 0.0149                     & 0.0426                     & 0.0085                     & 0.0177                      & 0.0636                     & 0.1211                     & 0.0419                     & 0.0602                      & 0.0687                     & 0.1307                     & 0.0562                     & 0.0793                      \\ \cline{2-14} 
                              & VAE                      & 0.0103                     & 0.0316                     & 0.0060                     & 0.0133                      & 0.0611                     & 0.1157                     & 0.0409                     & 0.0584                      & 0.0688                     & 0.1302                     & 0.0574                     & 0.0801                      \\
                              & VAE-AUR                  & 0.0196                     & 0.0480                     & 0.0123                     & 0.0218                      & 0.0648                     & 0.1218                     & 0.0430                     & 0.0613                      & 0.0698                     & 0.1328                     & 0.0575                     & 0.0809                      \\ \hline
                              & MF                       & 0.0036                     & 0.0098                     & 0.0022                     & 0.0043                      & 0.0374                     & 0.0679                     & 0.0251                     & 0.0349                      & 0.0412                     & 0.0777                     & 0.0325                     & 0.0461                      \\
                              & MF-AUR                   & 0.0235                     & 0.0448                     & 0.01524                    & 0.0222                      & 0.0637                     & 0.1060                     & 0.0439                     & 0.0575                      & 0.0550                     & 0.1003                     & 0.0434                     & 0.0602                      \\ \cline{2-14} 
Amazon                        & LGCN                     & 0.0072                     & 0.0190                     & 0.0043                     & 0.0083                      & 0.0381                     & 0.0708                     & 0.0254                     & 0.0359                      & 0.0436                     & 0.0820                     & 0.0347                     & 0.0489                      \\
                              & LGCN-AUR                 & 0.0229                     & 0.0468                     & 0.0145                     & 0.0224                      & 0.0525                     & 0.0918                     & 0.0356                     & 0.0483                      & 0.0511                     & 0.0936                     & 0.0399                     & 0.0557                      \\ \cline{2-14} 
                              & VAE                      & 0.0114                     & 0.0293                     & 0.0067                     & 0.0126                      & 0.0557                     & 0.0955                     & 0.0378                     & 0.0508                      & 0.0532                     & 0.0982                     & 0.0418                     & 0.0586                      \\
                              & VAE-AUR                  & 0.0285                     & 0.0543                     & 0.0187                     & 0.0272                      & 0.0607                     & 0.1031                     & 0.0416                     & 0.0553                      & 0.0581                     & 0.1053                     & 0.0458                     & 0.0635                      \\ \hline
\multicolumn{1}{|l|}{}        & MF                       & \multicolumn{1}{l}{0.0729} & \multicolumn{1}{l}{0.1419} & \multicolumn{1}{l}{0.0335} & \multicolumn{1}{l|}{0.0484} & \multicolumn{1}{l}{0.0852} & \multicolumn{1}{l}{0.1684} & \multicolumn{1}{l}{0.0400} & \multicolumn{1}{l|}{0.0574} & \multicolumn{1}{l}{0.0884} & \multicolumn{1}{l}{0.1652} & \multicolumn{1}{l}{0.0406} & \multicolumn{1}{l|}{0.0568} \\
\multicolumn{1}{|l|}{}        & MF-AUR                   & \multicolumn{1}{l}{0.1084} & \multicolumn{1}{l}{0.2187} & \multicolumn{1}{l}{0.0574} & \multicolumn{1}{l|}{0.0929} & \multicolumn{1}{l}{0.1213} & \multicolumn{1}{l}{0.2452} & \multicolumn{1}{l}{0.0626} & \multicolumn{1}{l|}{0.1000} & \multicolumn{1}{l}{0.1187} & \multicolumn{1}{l}{0.2400} & \multicolumn{1}{l}{0.0613} & \multicolumn{1}{l|}{0.0981} \\ \cline{2-14} 
\multicolumn{1}{|l|}{Adressa} & LGCN                     & \multicolumn{1}{l}{0.0716} & \multicolumn{1}{l}{0.1432} & \multicolumn{1}{l}{0.0452} & \multicolumn{1}{l|}{0.0477} & \multicolumn{1}{l}{0.1026} & \multicolumn{1}{l}{0.1910} & \multicolumn{1}{l}{0.0465} & \multicolumn{1}{l|}{0.0652} & \multicolumn{1}{l}{0.0994} & \multicolumn{1}{l}{0.1781} & \multicolumn{1}{l}{0.0452} & \multicolumn{1}{l|}{0.0619} \\
\multicolumn{1}{|l|}{}        & LGCN-AUR                 & \multicolumn{1}{l}{0.1265} & \multicolumn{1}{l}{0.1897} & \multicolumn{1}{l}{0.0665} & \multicolumn{1}{l|}{0.0800} & \multicolumn{1}{l}{0.1697} & \multicolumn{1}{l}{0.2536} & \multicolumn{1}{l}{0.0858} & \multicolumn{1}{l|}{0.1032} & \multicolumn{1}{l}{0.1529} & \multicolumn{1}{l}{0.2387} & \multicolumn{1}{l}{0.0761} & \multicolumn{1}{l|}{0.0948} \\ \cline{2-14} 
\multicolumn{1}{|l|}{}        & VAE                      & \multicolumn{1}{l}{0.0781} & \multicolumn{1}{l}{0.1490} & \multicolumn{1}{l}{0.0381} & \multicolumn{1}{l|}{0.0529} & \multicolumn{1}{l}{0.1071} & \multicolumn{1}{l}{0.1916} & \multicolumn{1}{l}{0.0510} & \multicolumn{1}{l|}{0.0690} & \multicolumn{1}{l}{0.0948} & \multicolumn{1}{l}{0.1761} & \multicolumn{1}{l}{0.0452} & \multicolumn{1}{l|}{0.0626} \\
\multicolumn{1}{|l|}{}        & VAE-AUR                  & \multicolumn{1}{l}{0.1335} & \multicolumn{1}{l}{0.2058} & \multicolumn{1}{l}{0.0677} & \multicolumn{1}{l|}{0.0832} & \multicolumn{1}{l}{0.1729} & \multicolumn{1}{l}{0.2645} & \multicolumn{1}{l}{0.0871} & \multicolumn{1}{l|}{0.1071} & \multicolumn{1}{l}{0.1471} & \multicolumn{1}{l}{0.2368} & \multicolumn{1}{l}{0.0729} & \multicolumn{1}{l|}{0.0923} \\ \hline
\multicolumn{1}{|l|}{}        & MF                       & \multicolumn{1}{l}{0.0001} & \multicolumn{1}{l}{0.0010} & \multicolumn{1}{l}{0.0001} & \multicolumn{1}{l|}{0.0005} & \multicolumn{1}{l}{0.0208} & \multicolumn{1}{l}{0.0448} & \multicolumn{1}{l}{0.0027} & \multicolumn{1}{l|}{0.0204} & \multicolumn{1}{l}{0.0039} & \multicolumn{1}{l}{0.0090} & \multicolumn{1}{l}{0.0040} & \multicolumn{1}{l|}{0.0055} \\
\multicolumn{1}{|l|}{}        & MF-AUR                   & \multicolumn{1}{l}{0.0429} & \multicolumn{1}{l}{0.0733} & \multicolumn{1}{l}{0.0280} & \multicolumn{1}{l|}{0.0364} & \multicolumn{1}{l}{0.0515} & \multicolumn{1}{l}{0.0947} & \multicolumn{1}{l}{0.0326} & \multicolumn{1}{l|}{0.0443} & \multicolumn{1}{l}{0.0415} & \multicolumn{1}{l}{0.0713} & \multicolumn{1}{l}{0.0285} & \multicolumn{1}{l|}{0.0369} \\ \cline{2-14} 
\multicolumn{1}{|l|}{ML-10M}  & LGCN                     & \multicolumn{1}{l}{0.0003} & \multicolumn{1}{l}{0.0010} & \multicolumn{1}{l}{0.0001} & \multicolumn{1}{l|}{0.0005} & \multicolumn{1}{l}{0.0146} & \multicolumn{1}{l}{0.0314} & \multicolumn{1}{l}{0.0019} & \multicolumn{1}{l|}{0.0145} & \multicolumn{1}{l}{0.0025} & \multicolumn{1}{l}{0.0060} & \multicolumn{1}{l}{0.0027} & \multicolumn{1}{l|}{0.0040} \\
\multicolumn{1}{|l|}{}        & LGCN-AUR                 & \multicolumn{1}{l}{0.0346} & \multicolumn{1}{l}{0.0648} & \multicolumn{1}{l}{0.0225} & \multicolumn{1}{l|}{0.0309} & \multicolumn{1}{l}{0.0455} & \multicolumn{1}{l}{0.0872} & \multicolumn{1}{l}{0.0284} & \multicolumn{1}{l|}{0.0404} & \multicolumn{1}{l}{0.0339} & \multicolumn{1}{l}{0.0638} & \multicolumn{1}{l}{0.0234} & \multicolumn{1}{l|}{0.0319} \\ \cline{2-14} 
\multicolumn{1}{|l|}{}        & VAE                      & \multicolumn{1}{l}{0.0056} & \multicolumn{1}{l}{0.0102} & \multicolumn{1}{l}{0.0032} & \multicolumn{1}{l|}{0.0075} & \multicolumn{1}{l}{0.0263} & \multicolumn{1}{l}{0.0566} & \multicolumn{1}{l}{0.0034} & \multicolumn{1}{l|}{0.0078} & \multicolumn{1}{l}{0.0094} & \multicolumn{1}{l}{0.0175} & \multicolumn{1}{l}{0.0075} & \multicolumn{1}{l|}{0.0149} \\
\multicolumn{1}{|l|}{}        & VAE-AUR                  & \multicolumn{1}{l}{0.0466} & \multicolumn{1}{l}{0.0841} & \multicolumn{1}{l}{0.0292} & \multicolumn{1}{l|}{0.0487} & \multicolumn{1}{l}{0.0509} & \multicolumn{1}{l}{0.0974} & \multicolumn{1}{l}{0.0311} & \multicolumn{1}{l|}{0.0694} & \multicolumn{1}{l}{0.0442} & \multicolumn{1}{l}{0.0812} & \multicolumn{1}{l}{0.0289} & \multicolumn{1}{l|}{0.0542} \\ \hline
\end{tabular}
}
\label{main result}
\end{table*}

\noindent \textbf{Evaluation Protocols. }
To measure the recommendation performance, we adopt two widely-used evaluation metrics: $Recall@K$ and $NDCG@K$ ($K= 20\,or\,50$). To evaluate both the overall recommendation and tail item recommendation performance, we define three evaluation protocols to rank items and compute the above metrics, as following: 

\begin{itemize} [leftmargin=*]
    \item \textbf{Overall} evaluation. In this protocol, we conduct \textbf{all-ranking}~\cite{PDA} -- all items that are not interacted by a user are the candidates and report the normal Recall and NDCG. Indeed, it measures the trade-off between recommendation of head items and tail items. 

    \item \textbf{Tail Absolute} evaluation. In this protocol, we focus on measuring the accuracy of recommending tail items when both head and tail items are candidates. We still conduct \textbf{all-ranking}, but just take the positive tail items in the testing set as  ground-truth. We adjust the evaluation metrics, 
    \eg $Recall@K=\frac{1}{|\mathcal{U}|}\sum_{u}\frac{|\mathcal{T}_u \cap \mathcal{R}^{K}_u|}{|\mathcal{T}_u|}$,
    where $T_{u} $ denotes the set of tail items that are interacted by the user $u$ in the testing set.  $\mathcal{R}^{K}_{u}$ denotes the top-$K$ recommended items among all candidates in the all-ranking protocol, and $|\cdot |$ denotes the size of the corresponding set. 
    
    \item \textbf{Tail Relative} evaluation. Refer to \cite{relative}, it also evaluates the tail recommendation performance but focuses on the ranking of tail items where all tail items not interacted by a user are the candidates. Correspondingly, we adjust the metric 
    $Recall@K=\frac{1}{|\mathcal{U}|}\sum_{u}\frac{|\mathcal{T}_u \cap {\mathcal{R}^{'}}^{K}_u|}{|\mathcal{T}_u|}$,
    where $ {\mathcal{R}^{'}}^{K}_u $ denotes the top-${K}$ recommended tail items.
\end{itemize}

\noindent \textbf{Compared Methods.} The proposed AUR generates recommendations with the consideration of uncertainties. To verify the effectiveness of the AUR framework, we implement AUR with three backbone models (MF, LGCN, and VAE), and compare AUR with corresponding backbone models that generate recommendations just based on expectations. 
\begin{itemize}[leftmargin=*]
    \item \textbf{MF}~\cite{2009Collaborative}. This refers to the Matrix Factorization method, which predicts user preference on an item as the inner product of user and item embedding. 
    
    \item \textbf{LGCN}~\cite{he2020lightgcn} refers to LightGCN, which takes the simplified graph convolution networks (GCN) to better capture the collaborative information in the user-item interaction graph. We fix the number of GCN layers to $3$ due to its good performance on both Yelp and Amazon~\cite{he2020lightgcn}.
    
    \item \textbf{VAE} \cite{liang2018variational} refers to MultiVAE. 
This is an item-based CF method built with the variational autoencoder (VAE).
    We set the network architecture as: $1024 \to 512 \to 1024$ and take Equation~\eqref{loss_backbone} 
    as the optimization objective. 
    
    \item \textbf{MF-AUR/LGCN-AUR/VAE-AUR}, refer to the AUR with MF, LGCN, and VAE as the expectation estimator, respectively. We set the network size of $q_i$ and $p_i$ in the uncertainty estimator as 1024 to fairly compare with VAE.    
\end{itemize}
To verify the ability of AUR to mitigate the popularity bias, we also compare MF-AUR with the following methods designed for addressing popularity bias applied to MF:
\begin{itemize}[leftmargin=*]
    \item \textbf{IPS}~\cite{liang2016causal}. This is a famous de-biasing method --- inverse propensity scoring,  which re-weights training samples to improve the tail recommendation.
    \item \textbf{MACR} \cite{wei2021model}. This is a SOTA method to eliminate the popularity bias in recommendation. It takes the counterfactual inference to remove the influence of popularity on recommendations, promoting the recommendation of tail items. We use the code released by the authors and tune hyperparameters in the same range as the MACR paper.
    \item \textbf{PopBias} \cite{popbias}. This is a method to reduce the model bias to achieve both high accuracy and high debias performance. It extends the existing BPR Loss with a new regularization term which regulates the score differences within positive and negative items. In our experiments, we choose the $ZeroSum$ around all variations proposed in the paper.
\end{itemize}

Moreover, we compare MF-AUR with two well-known probabilistic recommender models:
\begin{itemize}[leftmargin=*]
    \item \textbf{PMF} \cite{2007Probabilistic}, which assumes that latent features and labels obey different Gaussian distributions. The variance of each distribution is treated as a hyper-parameter.
    For fair comparison, we implement PMF with the same mini-batch strategy as AUR and tune all hyper-parameters in the ranges taken by the PMF paper. 
    
    \item \textbf{ExpoMF} \cite{2016Modeling}. 
     ExpoMF incorporates user exposure to items into collaborative filtering. It could avoid suppressing potential positive but mislabeled samples by decreasing the weights of negative samples exposed with lower probabilities. Regarding the exposure model of ExpoMF, we follow the 'per-item' strategy to implement exposure covariates, \ie encoding exposure via item popularity.  
    
\end{itemize}
\noindent \textbf{Hyper-parameter Settings. }
We optimize all models with Adam~\cite{2014Adam} and use the default learning rate $1e\text{-}4$ and default mini-batch size $32$. The $L_2$ regularization coefficient is searched in the range of $\{1e\text{-}4,1e\text{-}3,1e\text{-}2,1e\text{-}1\}$. For baselines with embedding layers (MF, LGCN, IPS, and MACR) and the expectation estimators of MF-AUR and LGCN-AUR, the embedding size is fixed to $128$. For MF, LGCN, VAE and all the expectation estimators of AUR, we take use of the same training strategy described in Section~\ref{expectation_training}, and the sampling rate $\mu$ of this strategy is searched in the range of $\{0.01,0.1,0.2,0.4\}$, and in most case the optimal value of $\mu$ is $0.1$.  For AUR, $\gamma$ is searched in the range of $\{1e\text{-}3,1e\text{-}2\}$, and the best values are $1e\text{-}3$ and $1e\text{-}2$ for Yelp and Amazon, respectively; $\beta$ is fixed to $1e\text{-}2$, and $\alpha$ is searched in the range of $\{1,2,3,4,5\}$ and the default choice is $1$; $\lambda$ is searched from $\{0,0.2,0.4,0.6,0.8,1.0\}$.

\subsection{Performance Comparison (RQ1)}
In this subsection, we first compare the proposed AUR with different expectation models (\ie MF, LGCN and VAE), then compare AUR with SOTA debiasing methods (IPS and MACR) and probabilistic approaches (PMF and ExpoMF). Lastly, we discuss the fairness of the comparisons.

\noindent\textbf{Compared with Expectation Models.}\label{main_exp}
We compare three expectation models (MF, LGCN, and VAE) with the corresponding versions of AUR (MF-AUR, LGCN-AUR and VAE-AUR). Table~\ref{main result} shows their recommendation performance on the four datasets under the three evaluation protocols. From the table, we have the following observations: 
\begin{itemize}[leftmargin=*]

    \item 
    In terms of the Overall evaluation, MF-AUR, LGCN-AUR, and VAE-AUR outperform their corresponding expectation models (MF, LGCN, and VAE) on all datasets except for Yelp. On Yelp, MF-AUR still outperforms MF, while LGCN-AUR and VAE-AUR show comparable performance to LGCN and VAE, respectively. These results confirm our theoretical analysis that incorporating uncertainty in generating recommendation scores can at least match the performance of expectation-based recommendations when considering the overall performance.

    \item 

    Regarding the Tail Absolute evaluation, all versions of the proposed AUR significantly outperform the corresponding expectation models. Specifically, the averaged relative improvements of AUR over the corresponding expectation-based models are $131.1\%$, $268.4\%$, $55.03\%$, and $687.17\%$ on Yelp, Amazon, Adressa, and ML-10M, respectively. These results indicate that AUR has a higher chance of retrieving relevant tail items, even when head items are also candidates, thus verifying that the recommendations are not biased toward head items. Given that the expectation-based model typically assigns low scores to tail samples that are mislabeled as negative, the substantial improvements in tail item recommendation performance achieved by AUR suggest that it can mitigate the impact of mislabeling by estimating the uncertainties.
    
    
    \item Regarding the Tail Relative evaluation, each version of AUR also outperforms its corresponding expectation model. The averaged RI is $12.2\%$, $37.0\%$, $50.91\%$, and $174.63\%$ for Yelp, Amazon, Adressa, and ML-10M, respectively.
    As only tail items are considered as ranking candidates in this evaluation protocol, these results suggest that AUR can better distinguish which tail items are more relevant to a user, achieving better Tail Relative performance. 
    Combined with the results of Tail Absolute, these findings demonstrate that AUR does not blindly provide higher recommendation opportunities for tail items, but rather identifies true relevant tail items, \ie potential positive tail items. 
    
    \item 
 Overall, regardless of the backbone recommender used (MF-based, graph-based, or VAE-based), AUR consistently demonstrates superior tail performance and comparable, or even better, overall performance in the top-$K$ recommendation. This confirms that AUR is model-agnostic and can be integrated into most existing models without the need for architecture changes or retraining, reducing the cost of deploying AUR in industry applications.
Additionally, achieving better tail performance through AUR is crucial for real-world applications. For instance, when item popularity shifts over time, promoting tail items that have the potential to become popular in the future could increase clicks and improve recommendation quality~\cite{PDA}.

    
    
    \item LGCN performs worse than VAE, which is different from the result in the original paper of LGCN~\cite{he2020lightgcn}. We think the reason is the difference of the training strategy. Note that the reported results of LGCN are better than the results reported in the LGCN paper, we believe the comparison is fair. Although LGCN outperforms MF, LGCN-AUR cannot outperform MF-AUR. This means although the AUR can improve different expectation model performance, the expectation estimator will also affect the performance of AUR, since the uncertainty estimation is based on results of the expectation estimator.

\end{itemize}

\begin{table}[]
\centering
\caption{The comparison between AUR, debiased models and probabilistic models regarding $Recall@50$ on Yelp and Amazon,  under the Overall, Tail Absolute, and Tail Relative evaluation protocols.}
\vspace{-3mm}
\small
\resizebox{0.45\textwidth}{!}{
\begin{tabular}{|c|c|ccc|}
\hline
Dataset                & Methods                      & Overall         & Tail Absolute   & Tail Relative   \\ \hline
                       & MF                           & 0.1215          & 0.0180          & 0.1010          \\
                       & IPS                          & 0.1103          & 0.0313          & 0.0916          \\
                       & MACR                         & 0.1193          & 0.0395          & 0.0950          \\
Yelp                   & \multicolumn{1}{l|}{PopBias} & 0.0810          & 0.0323          & 0.0763          \\
\multicolumn{1}{|l|}{} & ExpoMF                       & 0.1254          & 0.0315          & 0.1084          \\
\multicolumn{1}{|l|}{} & PMF                          & 0.1249          & 0.0226          & 0.1069          \\
                       & MF-AUR                       & \textbf{0.1316} & \textbf{0.0459} & \textbf{0.1230} \\ \hline
                       & MF                           & 0.0777          & 0.0098          & 0.0679          \\
                       & IPS                          & 0.0790          & 0.0298          & 0.0766          \\
                       & MACR                         & 0.0797          & 0.0183          & 0.0753          \\
Amazon                 & \multicolumn{1}{l|}{PopBias} & 0.0451          & 0.0174          & 0.0451          \\
\multicolumn{1}{|l|}{} & ExpoMF                       & 0.0815          & 0.0136          & 0.0718          \\
\multicolumn{1}{|l|}{} & PMF                          & 0.0781          & 0.0107          & 0.0682          \\
                       & MF-AUR                       & \textbf{0.1003} & \textbf{0.0448} & \textbf{0.1060} \\ \hline
\end{tabular}
}
\label{tab:sota-debiasing}
\vspace{-0.4cm}
\end{table}

\noindent\textbf{Compared with Debiased Models.}
We compare MF-AUR with IPS and MACR and present the results in Table~\ref{tab:sota-debiasing}. From the table, we can see that: 1) IPS, MACR and PopBias can only achieve obvious improvements on Tail Absolute evaluation, compared to MF. While on the Overall and Tail Relative evaluations, they can even be worse than the basic MF.  2) MF-AUR can beat IPS, MACR and PopBias on both the overall and tail evaluations, especially can achieve improvements in both Tail Absolute and Tail Relative evaluations. 
These results show that IPS, MACR and PopBias cannot recognize the relative relevance of tail items to a user more accurately than MF, \ie they indeed blindly promote the tail items, which is not an ideal solution for popularity bias~\cite{PDA}.
However, AUR can find truly relevant tail items from all tail candidates by identifying whether a tail item is indeed positive but mislabeled as negative.  


\noindent\textbf{Compared with Probabilistic Models.}
\label{exp_prob}
Considering that AUR is also a probabilistic modeling framework, we compare the AUR framework with existing probabilistic frameworks ExpoMF and PMF. The results are also shown in Table~\ref{tab:sota-debiasing}. From the table, we can draw the following conclusions: 1) Both ExpoMF and PMF outperform the basic MF model on all three evaluation protocols.
2) Compared with debiased models, ExpoMF and PMF achieve better overall performance but lower tail performance, 
while AUR brings improvements in both.
These results show that existing probabilistic approaches are better than non-probabilistic models. However, they fix the uncertainties (variance) as constants and take no account of uncertainty when generating recommendations.
They thus achieve much inferior performance than AUR, validating the superiority
of modeling the uncertainty in a learning manner and generating recommendation with the consideration of data uncertainty.

\begin{figure}[t]
	\centering
	\subfigbottomskip=1pt
	\subfigure{
		\includegraphics[width=0.8\columnwidth]{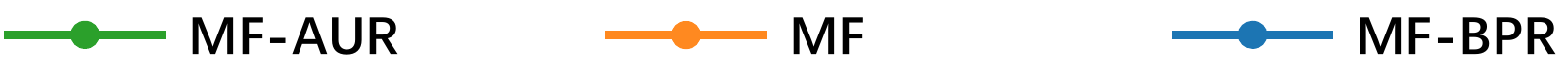}
	}
	\setcounter{subfigure}{0}
	\subfigure[Overall performance]{
		\includegraphics[width=0.44\columnwidth]{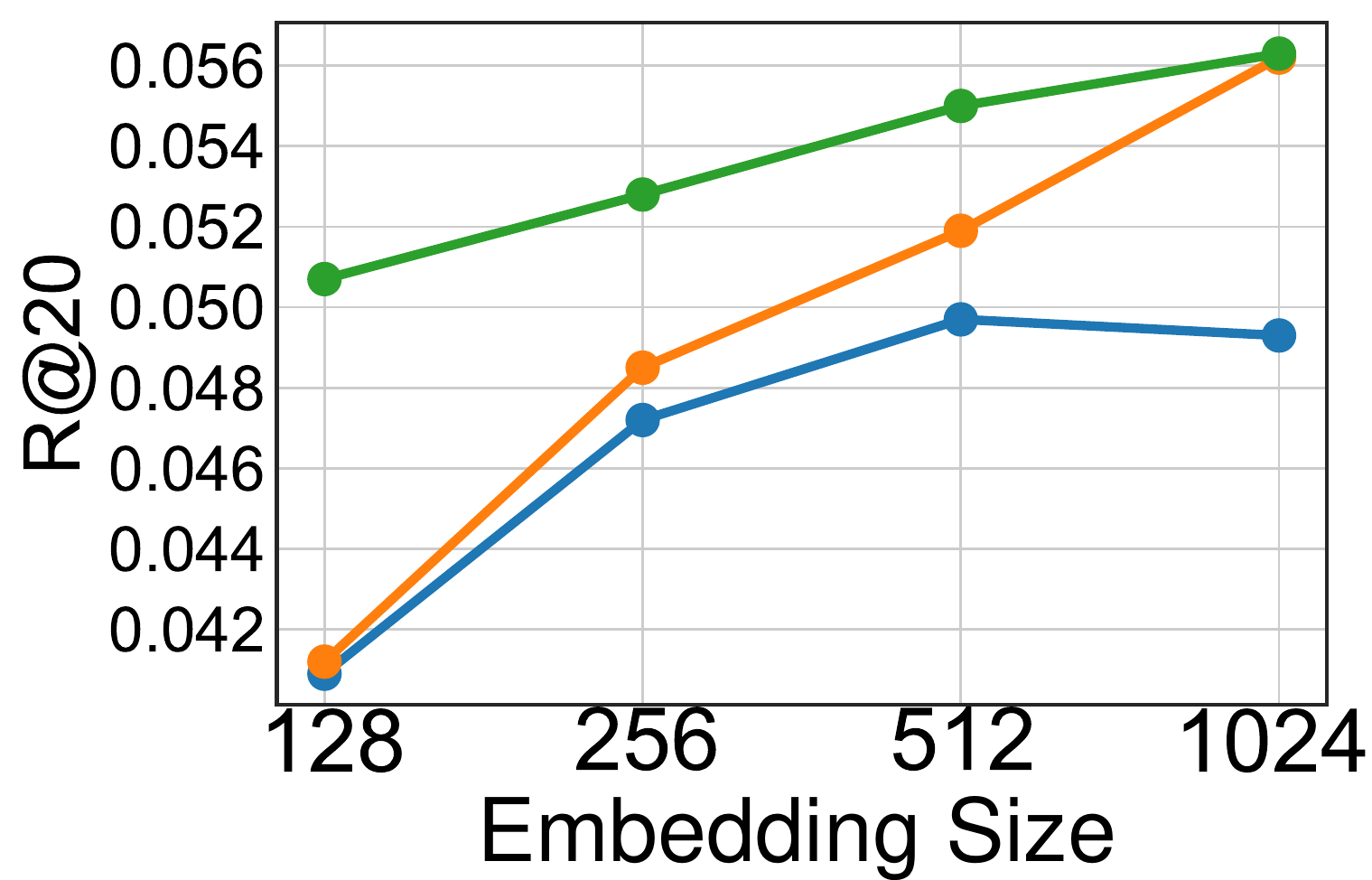}
	}
	\subfigure[Tail absolute performance]{
		\includegraphics[width=0.44\columnwidth]{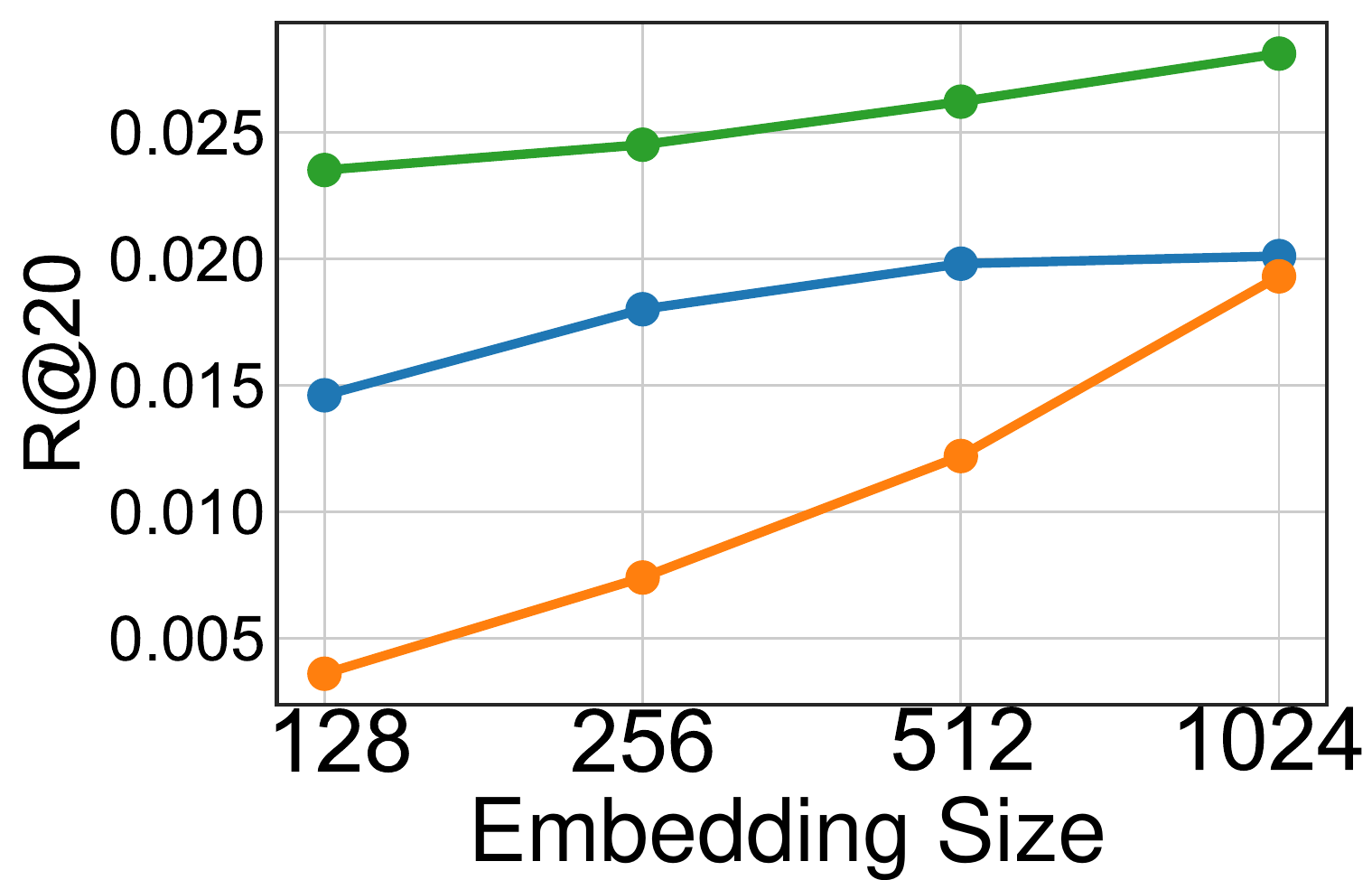}
	}
	\vspace{-2mm}
	\caption{The influence of training strategy and embedding size on MF, MF-BPR, and MF-AUR on Amazon.}
	\label{fair}
\end{figure}



\noindent\textbf{Comparison Fairness. }
 Recall that we take a new training strategy in Equation (4),  for MF, LGCN, and VAE, differing from typical training strategies, \eg optimizing BPR loss.
We question whether the training strategy decreases the effectiveness of these expectation-based models, leading to unfair comparison in Table~\ref{main result}. Thus we implement another MF with normal BPR loss, and denote it as MF-BPR. Besides, we also question whether the relative improvements of MF-AUR come from more model parameters, compared to the corresponding expectation-based model MF. Recall that the uncertainty estimator of AUR has an embedding size of $1024$, while the embedding size of MF is $128$. Thus we change the embedding size of MF, MF-BPR, expectation estimator (\ie MF) of MF-AUR to study its influence on model performance. The results are shown in Figure~\ref{fair}. From the table, we can find: 
\begin{table}[]
\centering
\small
\caption{The performance of different variants of AUR \wrt $Recall@50$.}
\vspace{-2mm}
\setlength{\tabcolsep}{2.5mm}{
\begin{tabular}{|c|cc|cc|}
\hline
\multirow{2}{*}{Methods} & \multicolumn{2}{c|}{Yelp} & \multicolumn{2}{c|}{Amazon} \\ \cline{2-5} 
                         & Overall      & Tail       & Overall       & Tail        \\ \hline
MF-AUR                   & 0.1316       & 0.0459     & 0.1003        & 0.0448      \\
MF-AUR-M                 & 0.1280       & 0.0343     & 0.0900        & 0.0280      \\
MF-AUR-J                 & 0.1349       & 0.0406     & 0.1097        & 0.0650      \\ \hline
LGCN-AUR                 & 0.1296       & 0.0447     & 0.0936        & 0.0468      \\
LGCN-AUR-M               & 0.1300       & 0.0422     & 0.0916        & 0.0353      \\
LGCN-AUR-J               & 0.1341       & 0.0451     & 0.1088        & 0.0659      \\ \hline
VAE-AUR                  & 0.1328       & 0.0518     & 0.1053        & 0.0543      \\
VAE-AUR-M                & 0.1340       & 0.0437     & 0.1029        & 0.0514      \\
VAE-AUR-J                & 0.1248       & 0.0490     & 0.1074        & 0.0656      \\ \hline
UE                       & 0.1111       & 0.0355     & 0.0948        & 0.0288      \\ \hline
\end{tabular}}
\label{variant}
\end{table}

\noindent $\bullet$ Regarding the training strategy, compared to MF that takes the proposed training strategy, MF-BPR that utilizes the BPR loss shows poor performance in the overall evaluation especially when the embedding size increases, and has relatively  better performance on Tail absolute evaluation. However, its superiority over MF on tail performance disappears when the embedding size is 1024. The tail and overall performances of MF-BPR are far lower than MF-AUR, which shows the advantages of the proposed training strategy.


\noindent $\bullet$ Regarding the embedding size, both MF and MF-AUR show increasing trends on overall and tail performance. 
MF achieves comparable results on overall performance to MF-AUR when the embedding size is 1024, while its tail performance is still far lower than MF-AUR. These results show that modifying the embedding size of MF does not change the above conclusions in this subsection.

\subsection{Studies on Model Designs (RQ2)}\label{ablation_study}
In this subsection, we investigate the influence of the key designs of AUR, including the training procedure, uncertainty estimator and the tail-controlling coefficient $\alpha$. 


\noindent \textbf{The Influence of Training Procedure and Uncertainty Estimator. } To study the influence of different training procedures and the design of the estimator, and verify the importance of the uncertainty estimation, we study three variants of AUR: 1) \textbf{AUR-J} takes a joint learning way to synchronously learn the expectation estimator and uncertainty estimator, \ie optimizing the objective in Equation~\eqref{loss_overall} to update them together. 2) \textbf{AUR-M} changes the uncertainty estimator to MLP, which takes the same multi-hot vectors as VAE as the input. For the MLP, there are two hidden layers both with the size of $1024$, and the activation function is ReLU. 3) \textbf{UE} 
estimates expectations using the same model and weighting strategy as uncertainty estimation, then generates recommendations based on the expectation. 
We compare these models with AUR on the four datasets under the three evaluation protocols. 
The results are shown in Table~\ref{variant}, where we omit the results on Tail Relative Evaluation that show similar trends to Tail Absolute Evaluation. 

From the table, we find, 1) compared to AUR, AUR-J shows better tail and overall performance on Amazon, but shows worse tail performance on Yelp, which is even worse than the expectation model (\textit{c.f.}, Table~\ref{main result}). This suggests selecting different training procedures for different datasets. In other words, different training procedures are suitable for different scenarios. However, considering the stability of AUR and its graftability to different backbone recommenders, we suggest taking the proposed two-step training (\ie sequential training) procedure of AUR. 
2) Although AUR-M takes a more complicated MLP as the uncertainty estimator, it performs worse than the original AUR in most cases. This demonstrates that our uncertainty estimator's architecture is effective. 
3) UE shows much worse tail performance on Yelp and Amazon, which verifies the importance of estimating the uncertainty for tail item recommendation. 

\begin{figure}[t]
\vspace{-2mm}
	\centering
	\subfigbottomskip=1pt
	\subfigure[$Recall@50$ on Yelp]{
		\includegraphics[width=0.44\columnwidth]{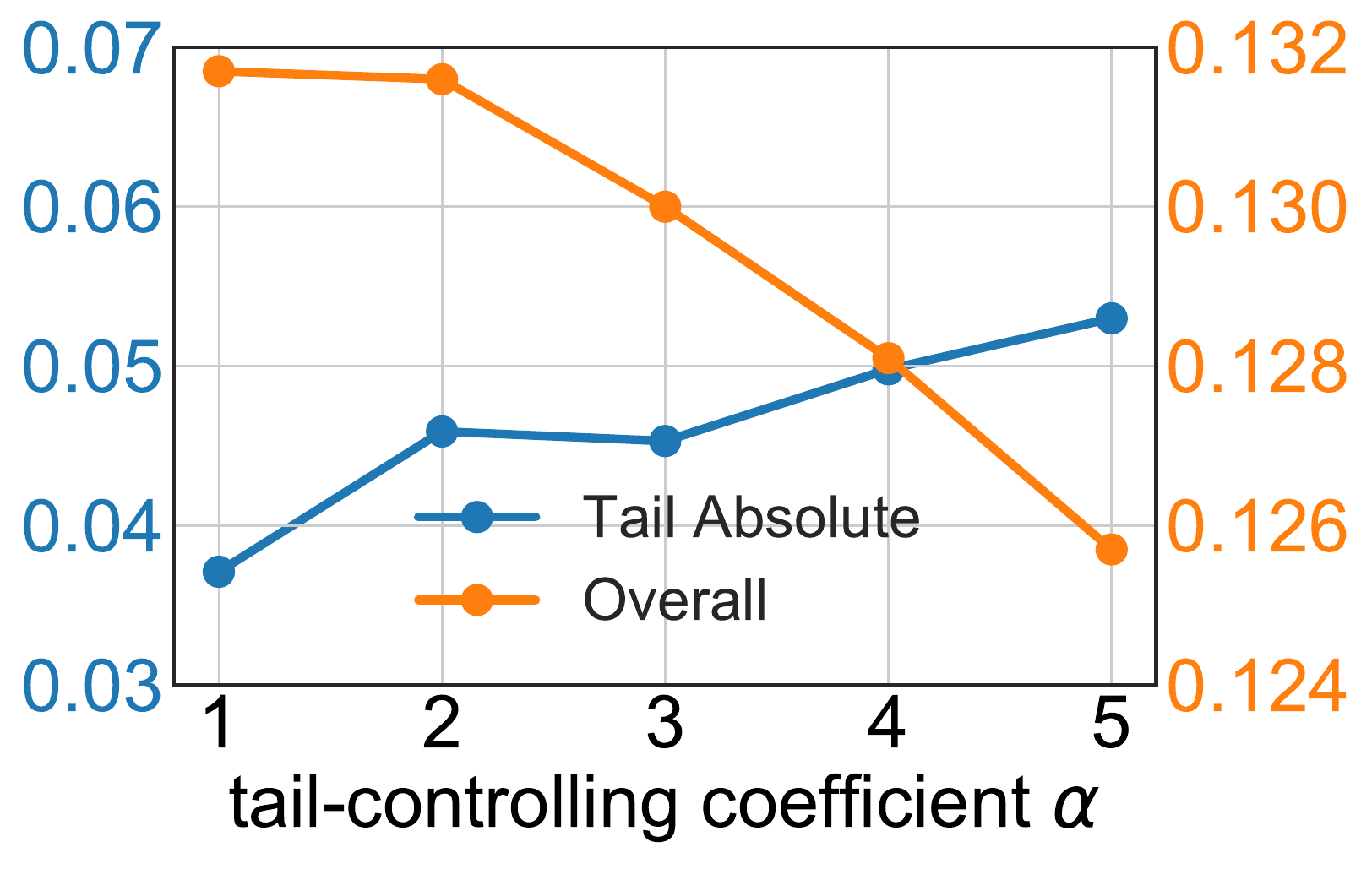}
	}
	\subfigure[$Recall@50$ on Amazon]{
		\includegraphics[width=0.44\columnwidth]{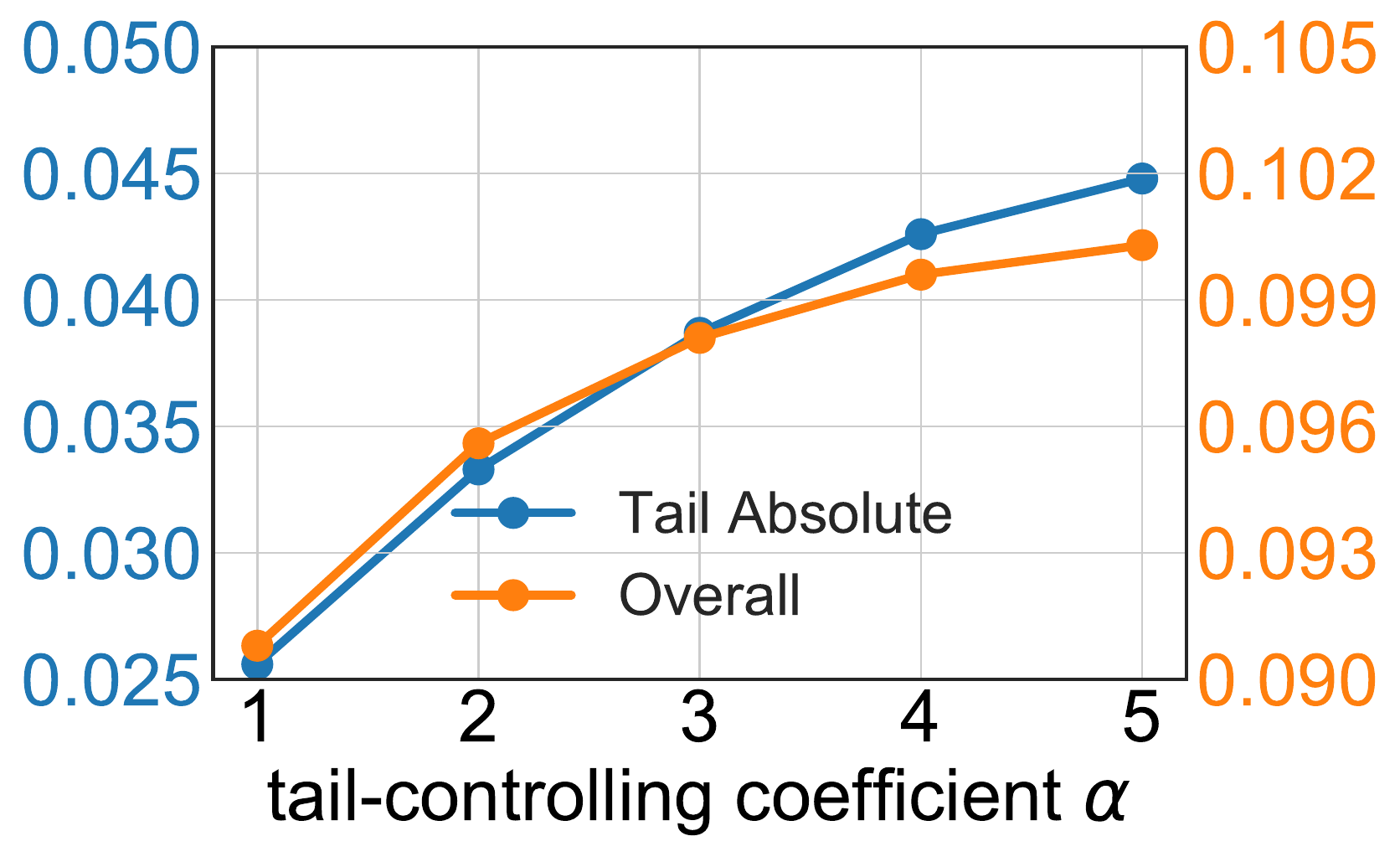}
	}
	\caption{The overall and tail performance of MF-AUR on Yelp and Amazon as changing the value of the
tail coefficient $\alpha$.}
	\label{hyper}
\end{figure}

\noindent \textbf{The Effect of Tail-controlling Coefficient $\alpha$. } Recall that the hyper-parameter $\alpha$ in Equation \eqref{weight}, \ie the tail-controlling coefficient, is to control the tail performance. To empirically study its influence on the model performance, we conduct an experiment that changes the $\alpha$ in the range of $\{1,2,3,4,5\}$ for MF-AUR, and draws the trends of overall and tail recommendation performance. The results \wrt $Recall@50$ are shown in Figure~\ref{hyper}. We omit the result on other metrics that exhibit the same trend. We find the tail performance keeps increasing in both Yelp and Amazon with $\alpha$ increasing, which supports the discussion that larger $\alpha$ indicates higher attention on the tail items in Section~\ref{sec:algo}.  
Regarding the overall performance, the result exhibits different trends in the four datasets, which may be attributed to the properties of the dataset. Note that the training and testing sets are chronologically split and the item popularity drifts largely in Amazon. This means the tail items of the training set may become head items in the testing set, thus dominating the overall performance, leading to both the increased overall and tail performances on Amazon. 
Therefore, it may be preferable to set a large value of $\alpha$ for a dataset with huge popularity drifts.

\vspace{-2mm}
\subsection{In-depth Analyses (RQ3)}
\begin{figure}[t]
	\centering
	\subfigbottomskip=1pt
	\subfigure{
		\includegraphics[width=0.8\columnwidth]{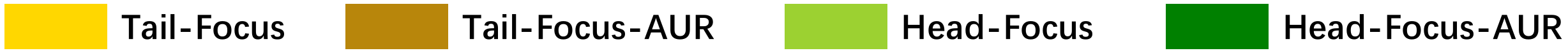}
	}
	\subfigure[Yelp]{
		\includegraphics[width=0.44\columnwidth]{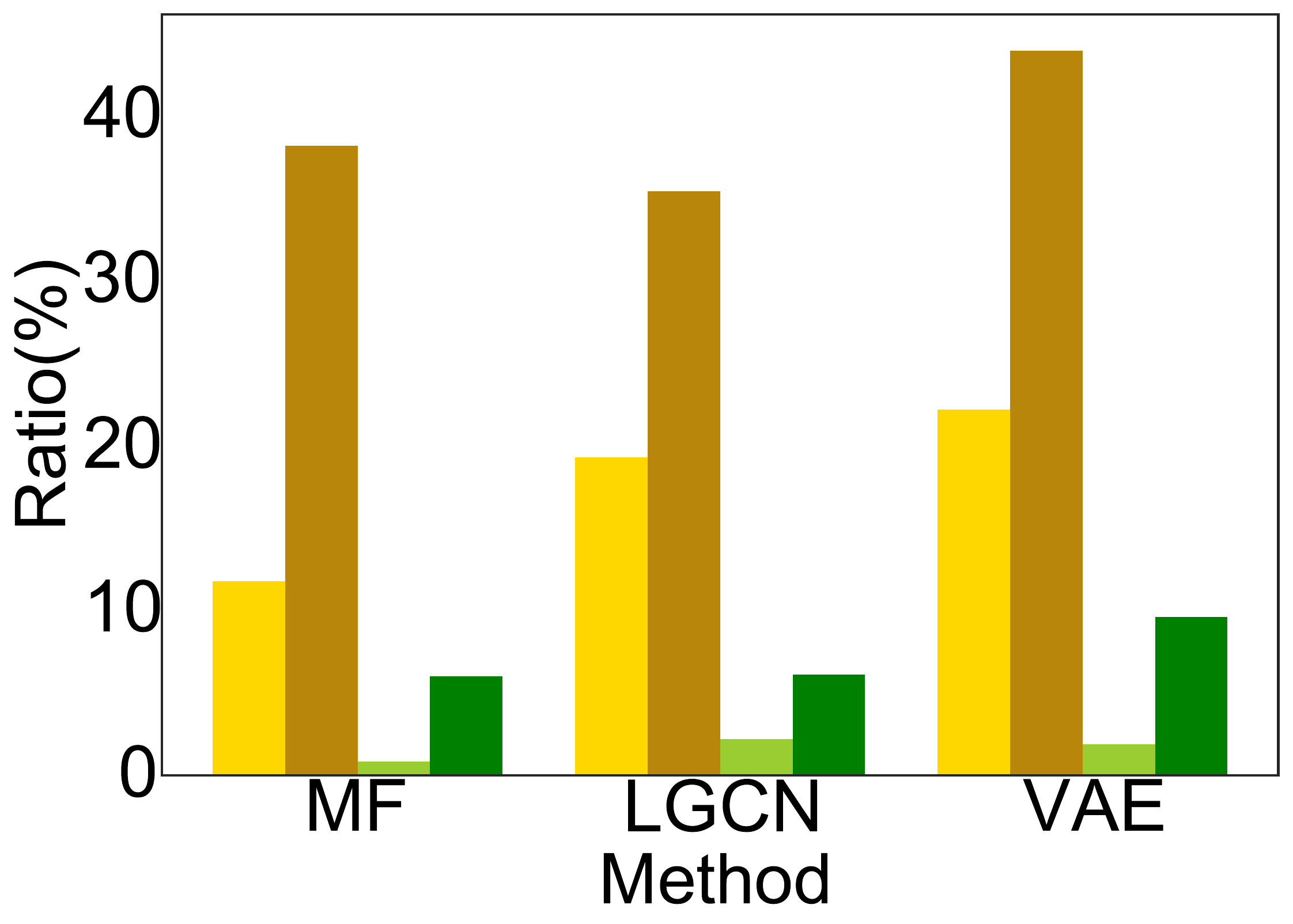}
	}
	\subfigure[Amazon]{
		\includegraphics[width=0.44\columnwidth]{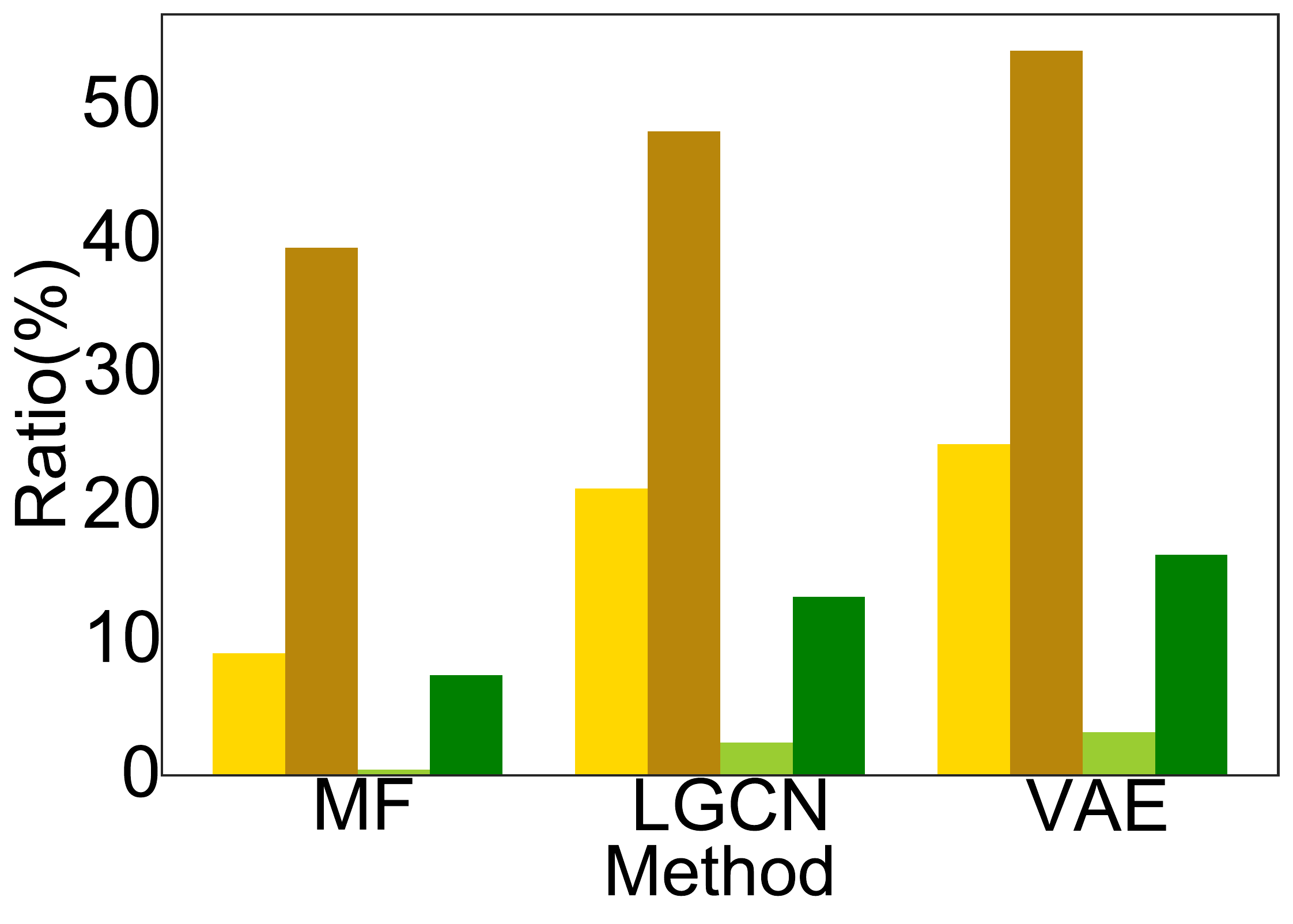}
	}
	\caption{Recommendation ratio of tail items in top 20 recommendations for tail-focus and head-focus users.}
	\label{tail proportion}
\end{figure}

\noindent \textbf{Calibrating Recommendation.}
Although most candidate items are tail items, traditional models prefer to recommend many head items to users even though certain users usually consume tail items. Thus we further study whether AUR calibrates the recommendation list regarding the tail and head items. We divide users into two groups: 1) tail-focus users whose at least $60\%$ of interacted items are tail items; 2) head-focus users who are not the tail-focus users. Then we count the proportion of tail items in the recommendation lists generated by AUR (MF-AUR,LGCN-AUR, and VAE-AUR) and expectation models (MF, LGCN, and VAE), for the tail-focus users and head-focus users, respectively. The result is shown in Figure~\ref{tail proportion}. We find: 1) AUR recommends more tail items to tail-focus users, and the recommendation ratio of tail items is greater than $35\%$ on Yelp and $40\%$ on Amazon, which is closer to the splitting ratio ($60\%$) compared to expectation models. 2) For head-focus users, expectation models almost do not recommend tail items, while AUR can recommend more tail items. 
The results  indicate that the ratios of tail items recommended by AUR are more close to that computed on the historical interactions. These findings verify that AUR can calibrate the recommendation ratios of tail items and head items, \ie recommending more suitable ratios of tail items to different types of users, compared with expectation models.

\begin{table}[]
\centering
\vspace{-2mm}
\caption{The averaged minimal length of recommendation list to cover all testing interactions over all users.}
\vspace{-3mm}
\resizebox{0.5\textwidth}{!}{
\begin{tabular}{|c|cccccc|}
\hline
Methods & MF     & MF-AUR & LGCN   & LGCN-AUR & VAE    & VAE-AUR \\ \hline
Yelp    & 18,823 & 15,546 & 16,738 & 10,894   & 21,041 & 12,387  \\ \hline
Amazon  & 42,255 & 41,184 & 39,453 & 28,824   & 50,684 & 41,831  \\ \hline
\end{tabular}
}
\label{worst_case}
\end{table}

\noindent \textbf{Worst Case Analysis. }
We then investigate the effect of AUR under the worst case by counting the number of recommendations needed to recall all relevant items in the testing set, \ie studying the minimal length of the recommendation list to cover all interacted items in the testing set. We average the required length over users for AUR and expectation model, where smaller value indicates better performance.
We summarize the results in Table~\ref{worst_case}. Compared with the expectation model, the corresponding AUR requires shorter recommendation lists to cover all positive testing interactions, which indicates the better robustness of making recommendation with uncertainty. Moreover, it reflects that generating recommendations with uncertainty taken into account can result in more item consumption at a lower cost. Meanwhile, LGCN-AUR achieves the best result on both Yelp and Amazon, which suggests selecting LGCN as the backbone expectation estimator of AUR when all relevant items are needed to be recommended.

\begin{table}[t]
\centering
\caption{The comparison of theoretical and empirical time complexity between MF, LGCN, VAE, and AUR on Amazon. The actual time cost per epoch (s) is calculated by averaging 10 epochs, and the $95\%$ confidence interval is reported.}
\scriptsize
\resizebox{88mm}{8mm}{
\begin{tabular}{|c|ccc|}
\hline
Time Analysis             & MF                & LGCN               & VAE                \\ \hline
MACC (1e6)                & 11.7              & 83.3               & 189.2              \\
Backbone (s)              & $31.11_{\pm0.02}$ & $149.23_{\pm0.09}$ & $100.13_{\pm0.05}$ \\
Uncertainty Estimator (s) & $92.35_{\pm0.07}$ & $125.37_{\pm0.09}$ & $104.95_{\pm0.10}$ \\ \hline
\end{tabular}
}
\label{time cost table}
\end{table}
\noindent \textbf{Time Cost Analysis. }
In this work, our AUR framework introduces an additional uncertainty estimation module. It is necessary to analyze the time complexity of the backbone expectation estimator and our AUR framework. Focusing on uncertainty estimator, we set $\lambda$ as $1$. 
We theoretically calculate the Multiply-Accumulate Operations(MACC)~\cite{mcquillanfeasibility} of each backbone method of AUR. We also empirically record the averaged training time cost per epoch on the Amazon dataset, for all models run on the same GPU machine (CPU:Intel(R) Core(TM) i9-9900X;GPU:NVIDIA GeForce RTX 2080 Ti;RAM:128GB). The results are summarized in the Table~\ref{time cost table}. 
The actual time of \textbf{Backbone(s)} and \textbf{Uncertainty Estimators} represents the actual training time cost of backbone training process and uncertainty estimator training process described in Section \ref{sec:training}. 

From the table, we see that: 
1)For LGCN and VAE, the actual time cost of uncertainty estimator is at the same level as backbone models. 
For MF, the actual time cost of uncertainty estimator is longer than the backbone model.
Because the time cost of uncertainty estimator contains feed-forward time of backbone model and uncertainty estimator, the uncertainty estimator is not much more complicated than MF. And in the Section \ref{ablation_study}, the variety \textbf{UE} proves that the performance improvement is not due to the more complex model.
Considering that AUR can improve tail performance and overall performance significantly, such additional cost is worthwhile.
2) Although the MACC of VAE is great than LGCN, VAE shows less actual time cost. This is because we only calculate the cost of the aggregation from neighbor nodes when computing the MACC of aggregation operation in LGCN. However, LGCN indeed implements the aggregation via matrix multiplication, bringing additional time cost.
3) The actual time cost of LGCN-AUR is even lower than LGCN. This is because the actual time cost contains not only the feed-forward time cost but also the back-propagation time cost. The gradient calculation and update of LGCN are complex and costly. 
When training LGCN-AUR, the LGCN is fixed. Therefore, the actual time cost of LGCN-AUR could decrease. 

 

%% file: related.tex
\section{Related Work}
\textbf{Uncertainty Quantification and Application.} 
According to the source of uncertainty, there are two types of uncertainty: epistemic uncertainty and aleatoric Uncertainty~\cite{uncertainty_survey}. Aleatoric uncertainty is also known as data uncertainty, which originates from randomness in the data generation process~\cite{uncertainty_survey}. The data mislabeling can be detected and addressed with this uncertainty~\cite{2017What,wang2019aleatoric}. 
The uncertainty has also been used to enhance the reliability and robustness of predictions against possible noise and corruption.
However, most methods are not designed for recommender systems. The epistemic uncertainty, which quantifies how uncertain a model is~\cite{uncertainty_app}, is less related to us. It is usually used to detect the out-of-distribution samples in risk-sensitive fields~\cite{al2019deep,leibig2017leveraging}, such as automatic control~\cite{al2019deep} and disease detection~\cite{leibig2017leveraging}. There is also the work~\cite{huang2022uncertainty} focusing on the label noise problem by leveraging both epistemic uncertainty and aleatoric uncertainty.
It is worth mentioning that our AUR framework is inspired by~\cite{2017What}.  We similarly assume that the labels obey Gaussian Distribution and learn the variance in a similar way. However, there are still some essential differences. \cite{2017What} trains the uncertainty to obtain a better expectation model, while we make recommendations based on the value of uncertainty. And they do not focus on
the improvement of tail prediction.

There are some efforts to model and utilize the uncertainty in recommendation~\cite{zeldes2017deep,wang2019bayesian,wang2020m2grl,liu2021personalised}.  
 The first line of research focuses on epistemic uncertainty~\cite{embedding_uncertainty,jiang2020convolutional,meta}. \cite{embedding_uncertainty} models the uncertainty of user/item representation to expand users' and items' interaction space, aiming at alleviating the cold-start problem. \cite{jiang2020convolutional}  employs Gaussian embedding, \ie endowing the randomness to the embedding, to adaptively capture the uncertain preference exhibited by some users. \cite{meta} introduces epistemic uncertainty to generate diverse recommendations for cold-start users. Different from them, we aim at modeling the uncertainty of data instead of the model. Secondly,  \cite{wang2020m2grl,xu2022ukd} pay attention to modeling the uncertainty from the data perspective. 
 \cite{wang2020m2grl} considers the multi-task scenarios and takes the uncertainty to identify the importance of different tasks.
\cite{xu2022ukd} introduces aleatoric uncertainty to mitigate the inherent noise in pseudo labels when distilling the knowledge from the teacher model.
Different from them, we consider a single task and directly take use of uncertainties to generate recommendations. Besides, \cite{zeldes2017deep,wang2019bayesian} model both the  epistemic uncertainty and aleatoric uncertainty in reinforcement learning-based recommendation. Their goal is to do better exploration and exploitation trade-offs with both the uncertainty and model normal prediction. Differently, we focus on collaborative filtering and pay attention to the mislabeling issue instead of the exploration and exploitation trade-off. \cite{gaussian} introduces Gaussian Process to make more accurate predictions for explicit recommendation, while we focus on addressing the mislabeling issue for implicit recommendation.

\textbf{Popularity Bias in Recommendation.} 
Tail item recommendation performance is very related to the popularity bias problem~\cite{ChenDQ0XCLY21,LeePL21}, which has gotten huge attention in the recommendation community. To deal with the popularity bias problem, existing methods can be divided into four different categories: 1) IPS-based methods~\cite{2019Unbiased,schnabel2016recommendations,becausal} deal with the problem by re-weighting training samples with propensities. However, the propensities are hard to set properly, causing high variance problems~\cite{PDA}. 2) Utilizing uniform data~\cite{liu2020general,zheng2021disentangling,ChenDQ0XCLY21} to guide the model to be unbiased. However, uniform data is usually small and expensive to gain. 3) The methods based on post-hoc re-ranking or model regularization~\cite{abdollahpouri2019managing,zhu2021popularity} usually need to heuristically design the re-ranking or regularization polices. 4) Causal inference-based methods such as~\cite{PDA,wei2021model} first take prior knowledge to create a causal graph to model the recommendation process, and then take intervention or counterfactual inference to eliminate the popularity bias. While the prior knowledge is not easy to get. Representation balancing, a causally deconfounding technique, is also used to deal with the MNAR problem~\cite{becausal}.
Different from all these methods, we take uncertainty modeling to address the mislabeling issue to achieve better recommendation performance for tail items.

\textbf{Data mislabeling.} 
To solve the mislabeling issue, \ie the positive-unlabeled issue, there are two main strategies: re-weighting and sampling.  Reweghting-based methods try to solve the problem by identifying mislabeled samples and assigning lower/higher weights for them/others. \cite{steck2010training,pan2008one,2009Collaborative,gantner2012personalized} take heuristic strategies for reweighting, such as assigning higher weights for unlabeled but popular items~\cite{gantner2012personalized}. \cite{chen2020fast} proposes to learn these weights instead of heuristically designing.
To solve the mislabeling issue, sampling-based methods avoid selecting the potential positive but unlabeled samples. There are many sampling methods~\cite{2015Improving,2016Modeling,wang2020m2grl,hsieh2014pu,yu2020sampler}. \cite{2015Improving} proposes a neighborhood-aware sampling strategy -- avoiding sampling items interacted by neighbors.  \cite{2016Modeling} gives higher sampling probabilities for unlabeled samples that have higher exposure probabilities. While the exposure probabilities are hard to be estimated well, \cite{wang2020m2grl} resorts to knowledge graph to design a sampling strategy. Besides, \cite{hsieh2014pu,yu2020sampler} directly estimates the probability of mislabeling with a Bayesian framework, and samples negatives based on this probability. Different to the two types of method, we learn the uncertainty of samples, and take the uncertainty as the mislabeling probability. Moreover, we directly take the uncertainty to form recommendation scores instead of utilizing them to re-weight or sample data.

%% file: conclusion.tex
\vspace{-3mm}
\section{Conclusion and Future Work}
In this work, we revealed that the aleatoric uncertainty can measure the mislabeling and tackle the positive-unlabeled issue. In this light, we proposed the AUR framework to estimate and leverage the aleatoric uncertainty.
To justify the rationality and effectiveness of AUR, we gave a theoretical proof 
and conducted experiments on two real-world datasets with three different evaluation protocols. 
The results demonstrated that our method can achieve better recommendation performance on tail items without sacrificing head items.
%
This work is an innovative attempt to introduce aleatoric uncertainty into recommendation. 
In the future, we will extend our method to other training paradigms, such as BPR and BCE. For the training process, we aim to overcome the unstable issue to achieve end-to-end training.
We are also interested in designing more powerful framework to estimate the uncertainty.
In addition, we will investigate whether the aleatoric uncertainty can mitigate other biases such as exposure bias~\cite{2016Modeling} and confounding bias~\cite{DCR}. Besides, we will further explore how to design a more effective uncertainty estimator.